\pdfoutput=1
\documentclass[12pt,a4paper]{article}

\ifnum\pdfoutput=1\else
\PassOptionsToPackage{hypertex}{hyperref}
\PassOptionsToPackage{draft}{graphicx}
\usepackage{showkeys}
\fi

\usepackage{amsmath,amssymb}
\usepackage[bookmarks=true,hyperfigures=true]{hyperref}
\usepackage{graphicx}
\usepackage[nosort]{cite}
\usepackage[bulletsep]{collref}

\usepackage[a4paper,text={450pt,656pt},centering]{geometry}

\let\oldbfseries=\bfseries
\let\oldmdseries=\mdseries
\let\oldnormalfont=\normalfont
\renewcommand{\bfseries}{\oldbfseries\boldmath}
\renewcommand{\mdseries}{\oldmdseries\unboldmath}
\renewcommand{\normalfont}{\oldnormalfont\unboldmath}

\allowdisplaybreaks[3]

\numberwithin{equation}{section}

\usepackage[font=small,labelfont=bf,width=0.85\textwidth]{caption}

\DeclareMathSymbol{\Gamma}{\mathalpha}{letters}{"00}
\DeclareMathSymbol{\Delta}{\mathalpha}{letters}{"01}
\DeclareMathSymbol{\Theta}{\mathalpha}{letters}{"02}
\DeclareMathSymbol{\Lambda}{\mathalpha}{letters}{"03}
\DeclareMathSymbol{\Xi}{\mathalpha}{letters}{"04}
\DeclareMathSymbol{\Pi}{\mathalpha}{letters}{"05}
\DeclareMathSymbol{\Sigma}{\mathalpha}{letters}{"06}
\DeclareMathSymbol{\Upsilon}{\mathalpha}{letters}{"07}
\DeclareMathSymbol{\Phi}{\mathalpha}{letters}{"08}
\DeclareMathSymbol{\Psi}{\mathalpha}{letters}{"09}
\DeclareMathSymbol{\Omega}{\mathalpha}{letters}{"0A}

\ifx\hypersetup\sadfkjashdfkxja\newcommand\hypersetup[1]{}\newcommand{\texorpdfstring}[2]{#1}\fi

\hypersetup{plainpages=false}
\hypersetup{pdfpagemode=UseNone}
\hypersetup{bookmarksnumbered=true}
\hypersetup{pdfstartview=FitH}
\hypersetup{colorlinks=false}
\hypersetup{citebordercolor={.5 1 .5}}
\hypersetup{urlbordercolor={.5 1 1}}
\hypersetup{linkbordercolor={1 .7 .7}}

\newcommand{\hypref}[2]{\ifx\hyperref\asklfhas #2\else\hyperref[#1]{#2}\fi}
\newcommand{\namedref}[2]{\hypref{#2}{#1~\ref*{#2}}}

\newcommand{\figref}[1]{\namedref{Fig.}{#1}}
\newcommand{\secref}[1]{\namedref{Sec.}{#1}}
\newcommand{\appref}[1]{\namedref{App.}{#1}}
\newcommand{\tabref}[1]{\namedref{Tab.}{#1}}

\newcommand{\atopfrac}[2]{\genfrac{}{}{0pt}{}{#1}{#2}}
\newcommand{\sfrac}[2]{{\textstyle\frac{#1}{#2}}}
\newcommand{\half}{\sfrac{1}{2}}
\newcommand{\quarter}{\sfrac{1}{4}}

\newcommand{\harm}[1]{h(#1)}

\newcommand{\alg}[1]{\mathfrak{#1}}
\newcommand{\grp}[1]{\mathrm{#1}}
\newcommand{\grSU}{\grp{SU}}

\newcommand{\grU}{\grp{U}}
\newcommand{\grSO}{\grp{SO}}
\newcommand{\grSp}{\grp{Sp}}
\newcommand{\grPSU}{\grp{PSU}}
\newcommand{\alU}{\alg{u}}
\newcommand{\alSU}{\alg{su}}
\newcommand{\alGL}{\alg{gl}}
\newcommand{\alSL}{\alg{sl}}
\newcommand{\alSO}{\alg{so}}
\newcommand{\alPSL}{\alg{psl}}

\newcommand{\alPSU}{\alg{psu}}

\newcommand{\fldA}{\mathcal{A}}
\newcommand{\fldF}{\mathcal{F}}
\newcommand{\order}[1]{\mathcal{O}(#1)}

\newcommand{\superN}{\mathcal{N}}
\newcommand{\gym}{g_{\scriptscriptstyle\mathrm{YM}}}
\newcommand{\gnorm}{g}

\newcommand{\Tr}{\mathop{\mathrm{Tr}}}

\newcommand{\indup}[1]{_{\mathrm{#1}}}

\newcommand{\Real}{\mathbb{R}}

\newcommand{\cder}{\mathcal{D}}

\newcommand{\Op}{\mathcal{O}}

\newcommand{\state}[1]{|#1\rangle}%
\newcommand{\vac}{\state{0}}%
\newcommand{\osca}{\mathbf{a}}%
\newcommand{\oscb}{\mathbf{b}}%
\newcommand{\oscc}{\mathbf{c}}%
\newcommand{\oscd}{\mathbf{d}}%
\newcommand{\oscA}{\mathbf{A}}%

\newcommand{\lrbrk}[1]{\left(#1\right)}
\newcommand{\bigbrk}[1]{\bigl(#1\bigr)}

\newcommand{\normord}[1]{\mathopen{:}#1\mathclose{:}}

\newcommand{\bigvev}[1]{\bigl\langle#1\bigr\rangle}
\newcommand{\bigcomm}[2]{\big[#1,#2\big]}
\newcommand{\comm}[2]{[#1,#2]}

\newcommand{\acomm}[2]{\{#1,#2\}}

\newcommand{\nln}{\nonumber\\}
\newcommand{\nl}{\nonumber\\&&\mathord{}}

\newcommand{\nle}{\nonumber\\&=&\mathrel{}}
\newcommand{\eq}{\mathrel{}&=&\mathrel{}}
\newenvironment{myeqnarray}{\arraycolsep0pt\begin{eqnarray}}{\end{eqnarray}\ignorespacesafterend}
\newenvironment{myeqnarray*}{\arraycolsep0pt\begin{eqnarray*}}{\end{eqnarray*}\ignorespacesafterend}

\def\[{\begin{equation}}
\def\]{\end{equation}}
\def\<{\begin{myeqnarray}}
\def\>{\end{myeqnarray}}

\ifx\href\asklfhas\newcommand{\href}[2]{#2}\fi
\newcommand{\arxivno}[1]{\href{http://arxiv.org/abs/#1}{#1}}

\begin{document}


\thispagestyle{empty}
\begin{flushright}\footnotesize
\texttt{\arxivno{hep-th/0307015}}\\
\texttt{AEI 2003-056}
\end{flushright}
\vspace{2cm}

\begin{center}
\hypersetup{pdftitle={The Complete One-Loop Dilatation Operator of N=4 Super Yang-Mills Theory}}%
\hypersetup{pdfkeywords={}}%
\hypersetup{pdfsubject={}}%
{\Large\textbf{The Complete One-Loop 
Dilatation Operator\\of 
$\superN=4$ Super Yang-Mills Theory}\par}
\vspace{2cm}

\textsc{Niklas Beisert}
\hypersetup{pdfauthor={Niklas Beisert}}%
\vspace{5mm}

\textit{Max-Planck-Institut f\"ur Gravitationsphysik\\
Albert-Einstein-Institut\\
Am M\"uhlenberg 1, D-14476 Golm, Germany}
\vspace{3mm}

\texttt{nbeisert@aei.mpg.de}\par\vspace{2cm}

\textbf{Abstract}\vspace{7mm}

\begin{minipage}{13.7cm}
We continue the analysis of \arxivno{hep-th/0303060}
in the one-loop sector and 
present the complete $\alPSU(2,2|4)$ dilatation operator of
$\superN=4$ Super Yang-Mills theory.
This operator generates the matrix of 
one-loop anomalous dimensions for all local operators
in the theory. 
Using an oscillator representation 
we show how to apply the dilatation generator to a
generic state.
By way of example, we determine the planar anomalous 
dimensions of all operators up to and including dimension $5.5$,
where we also find some evidence for integrability.
Finally, we investigate a number of subsectors
of $\superN=4$ SYM in which the dilatation operator simplifies. 
Among these we find the previously considered $\alSO(6)$ and $\alSU(2)$ subsectors, 
a $\alSU(2|4)$ subsector isomorphic to the BMN matrix model at one-loop,
a $\alSU(2|3)$ supersymmetric subsector of
nearly eighth-BPS states 
and, last but not least, a non-compact $\alSL(2)$ subsector
whose dilatation operator lifts uniquely to
the full theory.
\end{minipage}

\end{center}

\newpage


\vspace*{6pt}
\section{Introduction and conclusions}

The AdS/CFT correspondence
\cite{Maldacena:1998re,Witten:1998qj,Gubser:1998bc}
and some of its recent advances
involving plane waves and
spinning string solutions
\cite{Metsaev:2001bj,Berenstein:2002jq,Gubser:2002tv,Frolov:2002av}
have attracted attention
to the investigation of 
anomalous dimensions in $\superN=4$ Super Yang-Mills theory
\cite{Gliozzi:1977qd,Brink:1977bc}.
The virtue of this particular theory is that it has a conformal phase which
is not spoiled by quantum effects
\cite{Sohnius:1981sn,Howe:1984sr,Brink:1983pd}.
In fact, supersymmetry merges with conformal 
symmetry to the symmetry group $\grPSU(2,2|4)$. 
Composite local operators of the $\superN=4$ gauge theory
therefore arrange into multiplets or modules
of the algebra $\alPSL(4|4)$
\footnote{The conjugation property of the algebras will not be relevant here
and we will always consider the complex versions.
When we consider unitarity bounds, we will tacitly refer to the real
form $\alPSU(2,2|4)$.{}} \cite{Dobrev:1985qv}.
The modules are characterised by a set of numbers
which determine the transformation properties under 
$\alPSL(4|4)$. All of these numbers are (half-)integer valued, 
except the scaling dimension, which may
receive corrections due to quantum effects,
the so-called anomalous dimensions.

In the free theory, the action of the algebra 
on the space of local operators closes in a rather trivial manner.
When quantum corrections are switched on, however, the
transformation properties of operators change
in such a way that the closure of the algebra is preserved.
This puts tight constraints on consistent 
deformations of the algebra generators. 
In fact, these constraints have been employed to 
determine the first few anomalous dimensions 
\cite{Anselmi:1997mq,Anselmi:1998ms}.
A similar technique has later on been applied 
to obtain a remarkable all-loop result 
for an anomalous dimension \cite{Santambrogio:2002sb}.
There is some hope that consistency requirements uniquely
fix the deformations to a one-parameter family
representing different values of the coupling
constant $\gym$.

One way to obtain a consistent and interesting deformation
of generators clearly is to evaluate the 
deformations due to loop effects in $\superN=4$ SYM.
In a perturbative quantum field theory
this is however a non-trivial issue as
scaling dimensions are superficially fixed 
to their free theory values due to an absence
of a compensating scale. 
Nevertheless, this scale must be introduced 
in any attempt to regulate the divergencies
of quantum field theories.
In the regularised theory, the infinities
give rise to logarithms, which can be interpreted
as small shifts of scaling dimensions.
In that sense anomalous dimensions are 
intimately related to divergencies.
Although the divergencies are the source for anomalous dimensions,
they are also a major complication in higher loop computations.

The first few anomalous dimensions that have been 
calculated directly in $\superN=4$ SYM and up to two-loops 
\cite{D'Hoker:1998tz,Penati:1999ba,Penati:2000zv}
turned out to be zero:
The considered operators were BPS operators 
whose scaling dimensions are required to be protected in a
$\alPSL(4|4)$ symmetric theory,
thus confirming the superconformal nature of $\superN=4$ SYM.
Besides the BPS operators there are further operators
for which exceptional non-renormalisation theorems have been found
\cite{Arutyunov:2001mh}.
Non-vanishing anomalous dimensions up to two-loops have subsequently 
been calculated for the most simple, non-protected operator,
the Konishi operator \cite{Bianchi:1999ge,Bianchi:2000hn,Arutyunov:2001mh}.
The obtained anomalous dimension was in agreement with
the earlier, purely algebraic results. 
The complete tower of twist-two operators of which the Konishi
operator is the lowest example was investigated in 
\cite{Kotikov:2000pm,Kotikov:2001sc,Dolan:2001tt}.
Their one-loop scaling dimensions were found
to be $2n+(\gym^2 N/2\pi^2)h(2n)$, where $h(j)$ are the harmonic numbers%
\[\label{eq:IntroHarm}
\harm{j}:=\sum_{k=1}^j\frac{1}{k}=\Psi(j+1)-\Psi(1),
\]
which can also be expressed in terms of the digamma function 
$\Psi(x)=\partial\log\Gamma(x)/\partial x$.
Using methods of QCD deep inelastic scattering,
the DGLAP and BFKL equations, as well as 
by means of computer algebra systems developed 
for higher loop computations in the standard model,
this result has been generalised to two-loops 
\cite{Kotikov:2002ab,Kotikov:2003fb}.
Systematic means to obtain anomalous dimensions
involving four-point functions and superspace techniques 
were worked out in, e.g.\ \cite{Arutyunov:2000ku,Penati:2001sv}.
Finally, multi-trace operators and operators which 
mix in a non-trivial way have been 
investigated in \cite{Ryzhov:2001bp,Bianchi:2002rw,Arutyunov:2002rs}.

With the advent of the BMN correspondence 
\cite{Berenstein:2002jq} the attention
has been shifted away from lower dimensional operators
to operators with a large number of fields
\cite{Kristjansen:2002bb,Gross:2002su,Constable:2002hw,Beisert:2002bb,Constable:2002vq}.
There, the complications are mostly of combinatorial nature.
It was therefore desirable to develop efficient methods 
to determine anomalous dimensions without having to deal
with artefacts of the regularisation procedure.
In a purely algebraic way
the dilatation operator \cite{Beisert:2002ff,Beisert:2003tq}
generates the matrix of one-loop anomalous dimensions
for any set of operators which are made out of the
six scalars of $\superN=4$ SYM. 
What is more, the matrix of anomalous dimensions can be obtained 
exactly for all gauge groups and, 
in particular, for groups $\grSU(N)$ with finite $N$.
Even two or higher-loop calculations of anomalous dimensions,
which are generelly plagued by multiple divergencies,
are turned into a combinatorial excercise!
Using the dilatation generator techniques 
many of the earlier case-by-case studies 
of anomalous dimensions were easily confirmed.
They furthermore enabled a remarkable all-genus comparison 
between BMN gauge theory and
plane-wave string theory \cite{Spradlin:2003bw}.

Much progress has been made in recent months due
to integrable structures discovered in planar gauge theory and
free string theory.
Minahan and Zarembo realised that the planar 
one-loop dilatation operator for scalar operators 
is isomorphic to the Hamiltonian of a $\alSO(6)$ integrable spin chain
\cite{Minahan:2002ve}. 
Furthermore, there are indications that this remarkable result 
generalises to higher loops \cite{Beisert:2003tq}
potentially giving rise to a novel type of integrable model.
Using the Bethe ansatz \cite{Minahan:2002ve} is was possible to find 
states whose anomalous dimension can only be expressed in terms of 
elliptic functions \cite{Beisert:2003xu}. 
Astonishingly, corresponding 
string states of the same energy could be found 
using semi-classical methods \cite{Frolov:2003xy}.
This matching might be related to an integrable
structure found for free strings on $AdS_5\times S^5$
\cite{Bena:2003wd}.
It represents one of the strongest confirmations of
the AdS/CFT correspondence so far.

Most investigations of anomalous dimensions have focussed
on operators made out of scalars. For these the 
Feynman diagrams are comparatively easy to calculate.
In a few exceptions \cite{Gursoy:2002yy,Klose:2003tw,Chu:2003ji}
scalars with covariant derivatives 
have been considered. These calculations are notoriously difficult due to
the complex index structures and terms that are required
by conformal covariance of the correlators. 
Calculations involving fermions and field strengths have mostly been
avoided, however, with the aid of computer algebra they are feasible
\cite{Kotikov:2002ab,Kotikov:2003fb}. 
All in all not much is known about such operators, except maybe 
by means of supersymmetry arguments \cite{Beisert:2002tn}.
\bigskip

In the current work we would like to address the issue of 
generic operators of $\superN=4$ SYM and their one-loop anomalous dimensions.
We derive the \emph{complete} one-loop non-planar 
dilatation operator of $\superN=4$ SYM
\eqref{eq:Hfull}. 
For simplicity we will refer to this as the 
\emph{Hamiltonian}%
\footnote{On the background $\Real\times S^3$ it
is in fact the one-loop part of the Hamiltonian.}
$H$ and define
\[
D(\gnorm)=D_0+\frac{\gym^2 N}{8\pi^2} \, H+\order{\gnorm^3}.
\]
First of all we will investigate the form of generic local operators 
and how the Hamiltonian acts on them. 
Considering the structure of one-loop Feynman diagrams 
it is easy to see that the Hamiltonian acts on no more than
\emph{two} fields within the operator at the same time.
We could now go ahead and calculate these Feynman diagrams
for all combinations of two fields. 
Most of these calculations would turn out to be redundant
once superconformal symmetry is taken into account.
We therefore first investigate the independent 
coefficients of a generic superconformally invariant 
function of two fields.
These are in one-to-one correspondence to the 
irreducible multiplets in the tensor product 
of two field multiplets.
The irreducible multiplets can be distinguished by
their `total~spin'~$j$. 
We claim that the 
`Hamiltonian density' $H_{12}$ representing the Hamiltonian $H$ restricted 
to two fields $1,2$ is given by
\footnote{Similar Hamiltonians have appeared in the context of QCD 
energies (BFKL kernel) and one-loop anomalous dimensions (DGLAP kernel), 
see e.g.\ \cite{Faddeev:1995zg,Braun:1998id,Belitsky:2003ys}.
Note, however, that here the total spin operator $J_{12}$
corresponds to the quadratic Casimir of $\alPSL(4|4)$
instead of $\alSL(2)$.}
\[
H_{12}=2h(J_{12}), \qquad H=\sum_{k=1}^L H_{k,k+1}.
\]
where $J_{12}$ is an operator that measures the total spin $j$
of the two fields and where $h(j)$ is defined in \eqref{eq:IntroHarm}.
The above mentioned anomalous dimensions of twist-two operators 
are a trivial consequence of this.

In principle this result allows the computation of all
one-loop anomalous dimensions in $\superN=4$ SYM:
For every pair of fields within the operator, we 
decompose into components of definite total spin $j$. 
Each component we multiply by the harmonic number $2h(j)$
and add up all contributions. 
For non-planar corrections, in addition we have to compute the colour 
structure dictated by Feynman diagrammatics. 
In an explicit calculation this procedure has the drawback 
that the projection to total spin $j$ is rather involved.
Nevertheless, there is an alternative method to evaluate the 
action of the Hamiltonian density $2h(J_{12})$ which we will
name the \emph{harmonic action}. 
To describe the harmonic action we will represent
fields of $\superN=4$ in terms of excitations
of a \emph{supersymmetric harmonic oscillator},
see \cite{Gunaydin:1985fk}.
There is a natural way to do this
such that each excitation corresponds to a
spinor index of the field.
The harmonic action describes how to 
shuffle the oscillators (or spinor indices) between
the two involved fields in order to obtain $2h(J_{12})$.
We will demonstrate the action in terms of a simple example
and, as an application, we determine the spectrum of planar 
anomalous dimensions for single-trace operators of $\superN=4$ SYM
up to and including classical dimension $5.5$,
see \tabref{tab:anotab}.
Here, we also see some signs of integrability of the
complete planar dilatation generator. 
The issue of integrability will, however, not be discussed
in detail in this work; this is discussed in
\cite{Beisert:2003yb}.

Still, the harmonic action in its most general form requires some 
work to be applied. 
On the other hand, when we restrict to the $\alSO(6)$ subsector of operators
made out of scalar fields only, the action should simplify to 
the effective vertex of \cite{Beisert:2002bb}. 
This vertex consists of only two terms and is straightforwardly applied.
In the minimal $\alSU(2)$ subsector it was even possible to 
derive the two-loop contribution to the dilatation generator 
\cite{Beisert:2003tq}.
It would therefore be desirable to investigate further subsectors within 
which the action of the dilatation generator closes. 
Here, one should distinguish between
exactly closed subsectors (e.g.\ $\alSU(2)$) and
subsectors closed only at one-loop (e.g.\ $\alSO(6)$).
We will find a criterion for exactly closed subsectors
and determine all such sectors. 
Among these we find the subsectors relevant to
quarter-BPS and eighth-BPS 
operators. These are the above $\alSU(2)$ sector and a new 
$\alSU(2|3)$ sector. 
We also find a simple condition 
\eqref{eq:SecQuarterOp}
for quarter-BPS and eighth-BPS operators 
which we use to determine the 
lowest-lying eighth-BPS operator.
Among the one-loop subsectors we find the
above $\alSO(6)$ subsector, a non-compact $\alSO(4,2)$ brother
and a $\alSU(2|4)$ subsector in which the Hamiltonian agrees fully with
the one-loop Hamiltonian of the BMN matrix model
\cite{Berenstein:2002jq,Kim:2003rz}.

Probably the most interesting exactly closed subsector 
is the non-compact brother 
of the $\alSU(2)$ subsector, the $\alSL(2)$ subsector.
The main difference between the two is that 
in the $\alSU(2)$ sector there are only two
fields, $Z$ and $\phi$, while in the $\alSL(2)$ sector
there are infinitely many, $\cder^n Z$. At first sight these
sectors seem quite different, in terms of representations this
is however not the case. 
In the complex form both algebras are the same, 
the fields $Z,\phi$ transform in the
fundamental spin $\half$ representations, whereas
the fields $\cder^n Z$ transform in the 
spin $-\half$ representation.
In this new subsector we are able to compute the dilatation generator
\eqref{eq:BabyHam} by field theoretic means. 
To demonstrate the usefulness of this simplified Hamiltonian
we apply it to find a few anomalous dimensions,
see \tabref{tab:BabyEnergy}.
Finally, we will show
that the Hamiltonian lifts uniquely to the full
$\alPSL(4|4)$ Hamiltonian, i.e.\ it fixes all independent
coefficients of the most general $\alPSL(4|4)$ invariant form.
\medskip

This paper is organised as follows. 
We start by determining the most general 
form of the dilatation generator compatible
with $\alSL(4|4)$ invariance in \secref{sec:Form}. 
In \secref{sec:Baby} we investigate the
$\alSL(2)$ subsector, determine the Hamiltonian 
and show that it lifts uniquely to the full Hamiltonian.
We then introduce the 
oscillator representation of fields and 
work out the action of the Hamiltonian
in \secref{sec:Osc}.
\secref{sec:Sectors} contains an investigation 
of various subsectors within $\superN=4$ SYM.
We give an outlook in \secref{sec:Concl}.

\section{The form of the dilatation generator}
\label{sec:Form}

We start by investigating the general form of the one-loop
dilatation generator. We will see that
representation theory of the symmetry group as
well as Feynman diagrammatics puts tight constraints on the form.
What remains is a sequence of undetermined coefficients,
one for each value of `total spin', which will turn out to be the 
harmonic series. 

\paragraph{Letters and operators.}

Local operators $\Op$ are constructed as 
gauge invariant combinations of 
`letters' (irreducible covariant fields) $W_A$
\footnote{For convenience we will restrict to the gauge group $\grU(N)$. 
The letters $W_A$ are $N\times N$ matrices, which can be
parametrised by the generators $T^a$
in the fundamental representation of $\grU(N)$
as $W_A=T^a W_A^a$. Nevertheless, all results generalise straightforwardly 
to arbitrary gauge groups.}
\[\label{eq:WordOp}
\Op=\Tr W_\ast\cdots W_\ast\,\, \Tr W_\ast\cdots W_\ast \,\,\,\, \ldots 
\]
The letters of $\superN=4$ are the scalars, fermions and
field strengths as well as their covariant derivatives.
Here, traces and antisymmetries in derivative indices
are excluded; 
such fields can be reexpressed as reducible products of letters
via the equations of motion or Jacobi identities. 
The index $A$ in $W_A$ enumerates all such letters.

An alternative way of representing local operators is to use 
the state-operator map for $\superN=4$ SYM on $\Real\times S^3$,
which is conformally equivalent to flat $\Real^4$. 
When decomposing the fundamental fields into spherical harmonics on $S^3$
one gets as irreducible states precisely the same spectrum of letters $W_A$. 
In this picture the dilatation generator maps to the
generator of shifts along $\Real$, i.e.\ the Hamiltonian. 

\paragraph{Tree-level algebra.}

Under the superconformal algebra $\alPSL(4|4)$,
c.f. \appref{sec:alg},
consisting of the generators $J_0$, the letters transform among themselves
\[\label{eq:JonLetter}
J_0\, W_A=(J_0)^B{}_A W_B.
\]
Classically, the local operators $\Op$ 
transform in tensor product representations of \eqref{eq:JonLetter}.
A generator $J_0$ of $\alPSL(4|4)$ at tree-level 
can thus be written in terms of its action $(J_0)^B{}_A$ on a single letter $W_A$ as
\[\label{eq:Jacting}
J_0 \, W_A \cdots W_B  = (J_0)^C{}_A W_C \cdots W_B + 
\ldots + (J_0)^C{}_B W_A \cdots W_C.
\]
Using a notation for variation with respect to fields
\cite{Beisert:2003tq}
\[\label{eq:varydef}
\check W^A:=\frac{\delta}{\delta W_A}=T^a\frac{\delta}{\delta W^a_A}
\]
we can also write the tree-level generators as%
\footnote{Note the fusion and fission rules 
$\Tr X \check W^A \Tr Y W_B=\delta^A_B \Tr XY$,
$\Tr X \check W^A Y W_B=\delta^A_B \Tr X\Tr Y$.}
\[\label{eq:Jbare}
J_0=(J_0)^B{}_A  \Tr W_B \check W^A.
\]
The variation will pick any of the letters within an operator
and replace it by the transformed letter.
In particular the tree-level dilatation generator is
\[\label{eq:Dbare}
D_0=\sum\nolimits_{A} \dim(W_A)\, \Tr W_A \check W^A.
\]

\paragraph{Interacting algebra.}

When quantum corrections are turned on, 
the transformation properties of operators
change. In perturbation theory we will therefore
write the full generators $J$ as a series in the
coupling constant $\gnorm$,%
\footnote{The odd powers of $\gnorm$ are due to normalisation: 
Propagators are $\order{\gnorm^0}$, 
three-vertices are $\order{\gnorm}$.}
\[\label{eq:Jexpand}
J(g)=\sum_{k=0}^\infty \gnorm^k J_{k}.
\]
In this work we will concentrate on the first
correction to the dilatation generator, $D_2$.
It must be invariant under the tree-level 
algebra $J_0$. This follows from the
interacting algebra identity
\[\label{eq:DJcomm}
\comm{D(g)}{J(g)}=
\comm{D_0}{J_0+\gnorm J_1+\gnorm^2 J_2}
+\gnorm^2\comm{D_2}{J_0}
+\order{\gnorm^{3}}
=\dim(J) \,J(g)
\]
for every operator $J$ of the superconformal group.
In perturbation theory the bare dimension $D_0$ 
of all generators $J_k$ must be conserved 
\[\label{eq:BareDim}
\comm{D_0}{J_k}=\dim(J)\, J_k.\]
Eq.\ \eqref{eq:DJcomm} then implies for all $J_0$
\[\label{eq:D2inv}
\comm{J_0}{D_2}=0.\] 
In other words \emph{$D_2$ is
invariant under classical superconformal transformations}.
It will turn out that $D_2$ is invariant under 
another, nontrivial generator $B_0$
of the algebra $\alSL(4|4)=\alGL(1)\ltimes\alPSL(4|4)$,
see e.g.\ \cite{Intriligator:1998ig}.
We will refer to this additional $\alGL(1)$ hypercharge generator
as `chirality'. It is conserved at the one-loop level, but 
at higher loops it is broken due to the Konishi anomaly.
In what follows we will therefore consider 
the classical $\alSL(4|4)$ algebra of generators $J_0$;
the one-loop anomalous dilatation generator $D_2$ will
be considered an independent object, the Hamiltonian $H$,
\[\label{eq:JHdef}
J(g)=J+\order{\gnorm},\qquad 
D(g)=D+\frac{\gym^2N}{8\pi^2}\, H+\order{\gnorm^3},\qquad 
\comm{J}{H}=0.
\]

\paragraph{Generic form.}
\begin{figure}
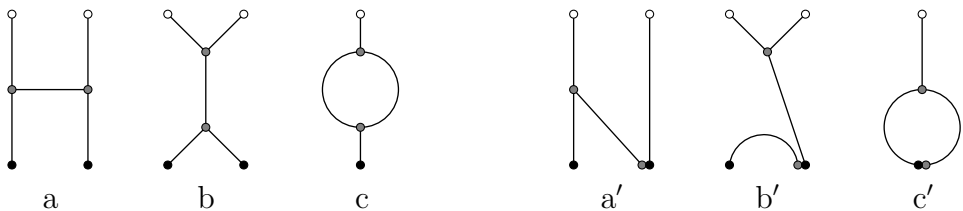
\centering
\parbox{1.5cm}{\centering\includegraphics{N4D2oneloopa.mps}\par a}
\quad
\parbox{1.5cm}{\centering\includegraphics{N4D2oneloopb.mps}\par b}
\quad
\parbox{1.5cm}{\centering\includegraphics{N4D2oneloopc.mps}\par c}
\qquad\qquad
\parbox{1.5cm}{\centering\includegraphics{N4D2oneloopap.mps}\par a$'$}
\quad
\parbox{1.5cm}{\centering\includegraphics{N4D2oneloopbp.mps}\par b$'$}
\quad
\parbox{1.5cm}{\centering\includegraphics{N4D2oneloopcp.mps}\par c$'$}
\caption{One-loop diagrams contributing to the anomalous dimension.
The lines correspond to any of the fundamental fields of the theory.}
\label{fig:diagrams}
\end{figure}%
The Hamiltonian 
has the following generic form
\<\label{eq:Hdiag}
H\eq
(C\indup{a})^{AB}_{CD} \,\normord{\Tr\comm{W_A}{\check W^C}\comm{W_B}{\check W^D}}
\nl
+(C\indup{b})^{AB}_{CD} \,\normord{\Tr\comm{W_A}{W_B}\comm{\check W^C}{\check W^D}}
\nl
+(C\indup{c})^{A}_{B} \,\normord{\Tr\comm{W_A}{T^a}\comm{T^a}{\check W^B}}.
\>
These terms correspond to the three basic types of Feynman diagrams
that arise at the one-loop level, see \figref{fig:diagrams}.
The diagrams a,b,c correspond to bulk interactions, 
where the covariant derivatives $\cder=\partial-i\gnorm \fldA$ within the fields
$W_A$ are reduced to a partial derivative.
The boundary interactions a$'$,b$'$,c$'$ arise 
when one covariant derivative emits a gauge field,
see also \cite{Gursoy:2002yy,Klose:2003tw}.
Algebraically they have the same form as their 
bulk counterparts. Diagrams with two emitted gauge fields 
have no logarithmic dependence and do not contribute to the
dilatation generator.
We can transform the term of type c by means of gauge invariance,
see \cite{Beisert:2003tq} for a detailed description of the procedure.
The generator of gauge transformations is
\[\label{eq:Gauge}
G^a=\Tr \comm{W_A}{\check W^A} T^a
\]
and it annihilates gauge invariant operators,
$G^a \mathrel{\widehat{=}} 0$.
Therefore we can write 
\[\label{eq:GaugeEquiv}
G^a \Tr \comm{W_A}{\check W^B} T^a  =
\normord{\Tr \comm{W_C}{\check W^C} \comm{W_A}{\check W^B} }
+\normord{\Tr \comm{W_A}{T^a} \comm{T^a}{\check W^B}}
\mathrel{\widehat{=}}0,
\]
which allows us to write the term of type c as a term of type a.
Furthermore the term of type b can be transformed by 
means of a Jacobi-identity. 
We combine all coefficients into a single one of type a
\[\label{eq:CoeffCombine}
C^{AB}_{CD}=
-N\bigbrk{(C\indup{a})^{AB}_{CD}
+(C\indup{b})^{AB}_{CD}
-(C\indup{b})^{AB}_{DC}
-\half\delta^A_C(C\indup{c})^{B}_{D}
-\half(C\indup{c})^{A}_{C}\delta^B_D}.
\]
The total Hamiltonian is
\[\label{eq:Hreduce}
H=
-N^{-1}C^{AB}_{CD} \,\normord{\Tr\comm{W_A}{\check W^C}\comm{W_B}{\check W^D}}.
\]
%

\paragraph{Symmetry.}

The combined coefficient $C^{AB}_{CD}$ must be invariant
under the tree-level superconformal algebra.
Its independent components can be obtained by
investigating the irreducible modules in the tensor product of
two multiplets of fields. 
We list all letters and their transformation 
properties in \tabref{tab:SYMfields}. 
\begin{table}\centering
$\begin{array}{|c|cccc|}\hline
\mathrm{field}       & \Delta_0 & \alSL(2)\times\alSL(2)&\alSL(4)&B\\\hline
\cder^k \fldF     & k+2 & [k+2,k\phantom{\mathord{}+0}] & [0,0,0] &+1\\
\cder^k \Psi      & k+\sfrac{3}{2} & [k+1,k\phantom{\mathord{}+0}] & [0,0,1] &+\sfrac{1}{2}\\
\cder^k \Phi      & k+1 & [k\phantom{\mathord{}+0},k\phantom{\mathord{}+0}] & [0,1,0] &\pm 0\\
\cder^k \bar \Psi & k+\sfrac{3}{2} & [k\phantom{\mathord{}+0},k+1] & [1,0,0] &-\sfrac{1}{2}\\
\cder^k \bar \fldF& k+2 & [k\phantom{\mathord{}+0},k+2] & [0,0,0] &-1\\\hline
\end{array}$
\caption{Components $W_A$ of the $\superN=4$ SYM field strength multiplet}
\label{tab:SYMfields}
\end{table}
We have split up the fermions and field strengths
into their chiral ($\Psi,\fldF$) and antichiral parts
($\bar\Psi,\bar \fldF$); the scalars ($\Phi$) are self-dual.
In the table
$\Delta_0$ is the bare dimension and refers the
$\alGL(1)$ of dilatations,
$\alSL(2)\times\alSL(2)$ is the Lorentz algebra
and $\alSL(4)$ is the flavour algebra.
These are the manifestly realised parts of the full
symmetry algebra $\alPSL(4|4)$.
We have also included the chirality $B$ of the
extended algebra $\alSL(4|4)$. 
In $\alSL(4|4)$ all letters $W_A$ combine into one
single, infinite-dimensional module
which we will denote by $V\indup{F}$.
The corresponding representation is also referred to as the `singleton'
representation.
Its primary weight 
\[\label{eq:Vfund}
w\indup{F}=[\Delta_0;s_1,s_2;q_1,p,q_2;B,L]=[1;0,0;0,1,0;0,1]\]
corresponds to one of the six scalars at unit dimension.
Here, $[q_1,p,q_2]$ are the Dynkin labels for a weight of $\alSL(4)$
and $[s_1,s_2]$ are \emph{twice} the spins of $\alSL(2)^2$.
The label $L$ refers to the number of fields. This will be of importance 
later on, here it equals one by definition.

The tensor product of two $V\indup{F}$ is given by 
\footnote{This statement can be proved by counting arguments using 
Polya theory (see e.g.\ \cite{Bianchi:2003wx}).}
\[\label{eq:VFTensor}
V\indup{F}\times V\indup{F}=
\sum_{j=0}^\infty V_{j}\]
where $V_j$ are the modules with primary weights
\<\label{eq:UIRs}
w_0\eq [2;0,0;0,2,0;0,2],\nln
w_1\eq [2;0,0;1,0,1;0,2],\nln
w_{j}\eq [j;j-2,j-2;0,0,0;0,2].
\>
Let $(P_j)^{AB}_{CD}$ project 
two field-strengths $W_A,W_B$
to the module $V_j$.
Then the most general invariant coefficients
can be written as
\footnote{In general, one might also consider Feynman diagrams in which 
two or three fields are transformed into one or vice versa.
These terms, however, cannot contribute to the leading order Hamiltonian,
which is $\alPSL(4|4)$ invariant,
because $V\indup{F}$ is not included in 
$V\indup{F}\times V\indup{F}$ and
$V\indup{F}\times V\indup{F}\times V\indup{F}$.}
\[\label{eq:D2Coeff}
C^{AB}_{CD}=\sum_{j=0}^\infty C_j\, (P_j)^{AB}_{CD}.
\]
In $\superN=4$ SYM we propose that the coefficients are given by 
the \emph{harmonic numbers} $\harm{j}$ or equivalently by the digamma function $\Psi$
\[\label{eq:HarmCoeff}
C_j=\harm{j}:=\sum_{k=1}^j\frac{1}{k}=\Psi(j+1)-\Psi(1).
\]
The one-loop dilatation generator 
of $\superN=4$ 
\eqref{eq:JHdef},\eqref{eq:Hreduce} 
can thus be written as
\[\label{eq:Hfull}
D(g)=D-\frac{\gym^2}{8\pi^2}\sum_{j=0}^\infty
\harm{j} \,(P_j)_{CD}^{AB} \,\normord{\Tr \comm{W_A}{\check W^C}\comm{W_B}{\check W^D}}
+\order{\gnorm^3}.
\]
This is the principal result of this work.
In \secref{sec:Osc} we will explain how
to compute the action of $D$ in practice.

\paragraph{Planar limit.}

At this point we can take the planar limit of
\eqref{eq:Hreduce}
and act on a single-trace operator of $L$ fields
\[\label{eq:HreducePlanar}
H=\sum_{k=1}^L H_{k,k+1},\qquad
H_{k,k+1}=2C_j P_{k,k+1,j},\]
where $P_{k,k+1,j}$ projects
the fields at positions $k,k+1$ to the module $V_j$.
We see that all coefficients $C_j$ can be read off
from this Hamiltonian.
Therefore the Hamiltonian density $H_{12}$ generalises uniquely
to the non-planar Hamiltonian $H$ in \eqref{eq:Hreduce},\eqref{eq:D2Coeff}.
In what follows 
we can safely restrict ourselves to the investigation of $H_{12}$.
The proposed Hamiltonian density according to 
\eqref{eq:Hfull} is
\[\label{eq:Hplanar}
H_{12}=\sum_{j=0}^\infty
2\harm{j} \,P_{12,j}.
\]
%

\paragraph{Spin functions.}

To simplify some expressions, we 
define functions $f(J_{12})$ of the total spin 
$J_{12}$ as 
\[\label{eq:FunctionOfJ}
f(J_{12})=\sum_{j=0}^\infty
f(j) \,P_{12,j}.
\]
In other words $f(J_{12})$ is a $\alPSL(4|4)$ invariant operator 
which acts on $V_j$ as 
\[\label{eq:FunctionOfJact}
f(J_{12})\, V_j= f(j)\, V_j.
\]
An arbitrary state with a dimension bounded by $\Delta_0$
belongs to the direct sum of modules $V_j$ with $j=0,\ldots,\Delta_0$.
The action of $f(J_{12})$ on such a state 
may be represented by a polynomial in the 
quadratic Casimir $J_{12}^2$, see \eqref{eq:gl44Casimir}. 
The eigenvalues of $J_{12}^2$,
where $J_{12}=J_1+J_2$ is the action of $\alPSL(4|4)$ on the tensor product 
$V\indup{F}\times V\indup{F}$,
are then given by
\[\label{eq:CasimirVj}
J_{12}^2 V_j=j(j+1) V_j.
\]
This can be proved using the methods of \secref{sec:OscRep}.
Note that all modules $V_j$ can be distinguished by the value of the 
quadratic Casimir.
We can therefore represent $f(J_{12})$ explicitly by
an interpolating polynomial in $J_{12}^2$
\[\label{eq:FuncOfJ}
f(J_{12})=
\sum_{j=0}^{\Delta_0}
f(j)
\prod_{\textstyle\atopfrac{k=0}{k\neq j}}^{\Delta_0}
\frac{J_{12}^2-k(k+1)}{j(j+1)-k(k+1)}.
\]
We note that this action also preserves chirality $B$, because
the quadratic Casimir $J_{12}^2$ does.
Therefore any $\alPSL(4|4)$ invariant function 
acting on $V\indup{F}\times V\indup{F}$ 
also conserves $\alSL(4|4)$. 
Clearly, this will not be the case for higher-loop
corrections to the dilatation generator which act on more
than two fields at the same time.
At higher loops the Konishi anomaly mixes operators of different
chirality. The same points also hold for the length $L$ of a state. 
Nevertheless it makes perfect sense to speak of the 
leading order chirality and length to describe a state. 
Mixing with states of different chiralities
or lengths is subleading, because the one-loop dilatation generator 
conserves these.
Using the short notation the Hamiltonian density becomes simply
\[\label{eq:HplanarShort}
H_{12}=2\harm{J_{12}}.
\]
%

\paragraph{Examples.}

A straightforward exercise is to determine the spectrum of
operators of length $L=2$. These so-called twist-two operators 
can be conveniently written as
\[\label{eq:Pomeron}
\Op_{j,AB}=(P_j)^{CD}_{AB} \Tr W_C W_D\sim P_{12,j}\Tr W_1 W_2.
\]
This operator vanishes for odd $j$ due to antisymmetry.
Using \eqref{eq:Hfull},\eqref{eq:HreducePlanar} or \eqref{eq:HplanarShort}
we find 
\[\label{eq:PomeronEnergy}
E=4h(j),\qquad \delta D=\frac{\gym^2 N}{8\pi^2}\,E=\frac{\gym^2 N}{2\pi^2}\,h(j)
\]
in agreement with the results of \cite{Kotikov:2000pm,Kotikov:2001sc,Dolan:2001tt}.
In fact the result of \cite{Kotikov:2000pm,Kotikov:2001sc,Dolan:2001tt} could be used to fix 
the even coefficients $C_{2k}$ \eqref{eq:HarmCoeff}.

\section{The non-compact \texorpdfstring{$\alSL(2)$}{sl(2)} closed subsector}
\label{sec:Baby}

In this section we will consider a closed subsector of 
states in $\superN=4$ SYM. 
We derive the relevant part of the 
Hamiltonian $H'$ and show that it uniquely lifts to the 
full Hamiltonian $H$ of $\superN=4$ SYM. 
Finally, we apply the Hamiltonian within this subsector 
and obtain a few anomalous dimensions.

\paragraph{Letters and operators.}

Let us consider single-trace operators in the planar limit
which saturate the bound $\Delta_0\geq L+n$, 
where $n$ and $L$ are the charges
with respect to rotations in the 
spacetime $12$-plane and flavour $56$-plane, respectively.
The operators consist only of the scalar field
$\Phi_{5+i6}=\Phi_5+i\Phi_6=\Phi_{34}$ and 
derivatives $\cder_{1+i2}=\cder_{1}+i\cder_{2}=\cder_{11}$ acting on it.%
\footnote{Depending on the situation, we take the freedom to use either spinor or vector index notation.}
A letter for the construction of operators
is thus 
\[\label{eq:BabyLetter}
(\osca^\dagger)^n \vac=\frac{1}{n!} (\cder_{1+i2})^n \Phi_{5+i6}.
\]
This subsector is closed, its operators do not mix
with any of the other operators in $\superN=4$ SYM.
This is similar to the subsector considered in 
\cite{Beisert:2003tq} where both charges belong to $\alSL(4)$. 
We have collected further interesting subsectors in 
\secref{sec:Sectors}.
The weight of a state with a total number of $n$
excitations (derivatives) is given by 
\[
w=[L+n;n,n;0,L,0;0,L].
\]
For $n\neq 0$ this weight is beyond the unitarity bounds 
and therefore it cannot be primary. 
The corresponding primary weight is
\[
w=[L+n-2;n-2,n-2;0,L-2,0;L].
\]
%

\paragraph{Symmetry.}

The subsector is invariant under an $\alSL(2)$ subalgebra
of the superconformal algebra
(note that $L^1{}_1=\dot L^1{}_1=\half D-\half L$ in this sector)
\<\label{eq:BabyAlg}
J'_+\eq P_{11}=P_{1+i2},
\nln
J'_-\eq K^{11}=K^{1+i2},
\nln
J'_3\eq\half D+\half \delta D+\half L^1{}_1+\half \dot L^1{}_1=D+\half \delta D-\half L.
\>
Here, the dilatation generator $D$ is part of the algebra.
At higher loops, one should keep in mind that only half of 
the anomalous piece appears. The $\alSL(2)$ subalgebra follows from
the relations in \appref{sec:alg}
\[\label{eq:BabyComm}
\comm{J'_+}{J'_-}=-2J'_3,\qquad \comm{J'_3}{J'_\pm}=\pm J'_\pm.
\]
It has the quadratic Casimir operator
\[\label{eq:BabyCasimir}
J^{\prime\,2}=J_3^{\prime\, 2}-\half\acomm{J'_+}{J'_-}.\]

The algebra may be represented by means of oscillators
$\osca$, $\osca^\dagger$
\[\label{eq:SL2alg}
J'_-= \osca,
\quad
J'_3= \half+\osca^\dagger \osca,
\quad
J'_+= \osca^\dagger+\osca^\dagger\osca^\dagger \osca.
\]
We define the canonical commutator as
\[
\comm{\osca}{\osca^\dagger}=1
\]
and assume that the vacuum state $\vac$ is annihilated by $\osca$.
Under this algebra the set of letters $(\osca^\dagger)^n\vac$ 
transforms in the infinite dimensional 
spin $j=-\half$ representation $V'\indup{F}$ with highest weight
\[\label{eq:BabyVf}
w'\indup{F}=[2j]=[-1].\]
The Hamiltonian density $H'_{12}$ acts on two fields $(\osca^\dagger_1)^k (\osca^\dagger_2)^{l}\state{00}$.
Of particular interest is therefore the tensor 
product of two $V'\indup{F}$, it splits into
modules of spin $-1-j$.
\[\label{eq:BabyVfSquare}
V'\indup{F}\times V'\indup{F}=\sum_{j=0}^\infty V'_j,
\qquad\mbox{with}\quad w'_j=[-2-2j].
\]
The algebra acting on this tensor product is given by the generators 
$J'_{12}=J'_1+J'_2$. 
The quadratic Casimir reads
\[\label{eq:BabyCasimirTwo}
J^{\prime\,2}_{12}=-(\osca_1^\dagger-\osca_2^\dagger)^2\osca_1\osca_2
+(\osca_1^\dagger-\osca_2^\dagger)(\osca_1-\osca_2).
\]
The highest weight state of $V'_j$,
which is annihilated by $J'_{12,-}$,
is 
\[\label{eq:BabyHWState}
\state{j}=(\osca^\dagger_1-\osca^\dagger_2)^j\,\state{00}.\]
the eigenvalue of the quadratic Casimir 
in that representation is 
\[\label{eq:BabyCasimirValue}
J_{12}^{\prime\, 2}\, V'_j= j(j+1) V'_j,
\]
which intriguingly matches the $\alPSL(4|4)$ counterpart 
\eqref{eq:CasimirVj}.

\paragraph{The Hamiltonian.}

Using point splitting regularisation we find the action of the 
Hamiltonian density
\[\label{eq:BabyHam}
H'_{12}\, (\osca^\dagger_1)^k (\osca^\dagger_2)^{n-k}\state{00}=
\sum_{k'=0}^n
\lrbrk{\delta_{k=k'}\bigbrk{\harm{k}+\harm{n-k}}-\frac{\delta_{k\neq k'}}{|k-k'|}}
(\osca^\dagger_1)^{k'} (\osca^\dagger_2)^{n-k'}\state{00},
\]
see \appref{sec:diagrams} for details.
It is straightforward to verify
that this $H'_{12}$ is invariant under the generators
$J'_{12}$.
In analogy to \eqref{eq:Hplanar} it is therefore clear that
we can write $H'_{12}$ as
\[\label{eq:BabyHamIndep}
H'_{12}=\sum_{j=0}^\infty 2C'_j\, P'_{12,j}.
\]
where $P'_{12,j}$ projects a state $(\osca^\dagger_1)^k (\osca^\dagger_2)^{n-k}\state{00}$
to the module $V'_j$.
The coefficients $C'_j$ are found by acting with $H'_{12}$ on 
the highest weight states $\state{j}$. Using the sum
\[\label{eq:BabySum}
\sum_{k=a+1}^n\frac{(-1)^{a+k+1}a!(n-a)!}{k!(n-k)!(k-a)}=\harm{n}-\harm{a}
\]
we can show that 
\[\label{eq:BabyHamAct}
2C'_j\,\state{j}=H'_{12}\,\state{j}=\harm{j}\,\state{j}.\]
Note that the Hamiltonian density \eqref{eq:BabyHamIndep} 
has precisely the right eigenvalues \eqref{eq:BabyHamAct} for total spin $-1-j$ 
in order to be integrable \cite{Faddeev:1996iy}. 
Thus we have found that the planar Hamiltonian in this subsector is 
the Hamiltonian of the XXX$_{-1/2}$ Heisenberg spin chain.
This is investigated in further detail in 
\cite{Beisert:2003yb}.

\paragraph{Lift to $\alPSL(4|4)$.}

The interesting feature of this subsector is 
that there is a one-to-one correspondence between
the modules $V_j$ \eqref{eq:UIRs} and $V'_j$.
The Hamiltonian in this subsector lifts to the full $\superN=4$ Hamiltonian!
Clearly, as $\alSL(2)$ is a subalgebra of
$\alPSL(4|4)$, the module $V'_j$ is 
a submodule of some $V_{j'}$. 
Here, it turns out, that $j'=j$
and
\[\label{eq:BabyRepLift}
V'_j\subset V_j.\]
In the $\alPSL(4|4)$ algebra the state $\state{j}$ has weight 
\[\label{eq:BabyWeight}
w_j'=[j+2;j,j;0,2,0;0,2].\]
For $j=0$ this is the primary weight of the 
current multiplet $w_0=[2;0,0;0,2,0;0,2]$. 
For ${j>0}$ the weight is beyond
the unitarity bounds and cannot 
be primary. The corresponding 
primary weights are $w_j=[j;j-2,j-2;0,0,0;0,2]$
or $w_1=[2;0,0;1,0,1;0,2]$, respectively.
These are exactly the primary weights of the modules
$V_j$ in \eqref{eq:UIRs}.
Using the fact that the two Hamiltonians must agree within 
the subsector we find
\[\label{eq:BabyLift}
2C_j \,\state{j}=H_{12}\, \state{j}=H'_{12}\,\state{j}
=2\harm{j}\, \state{j}.
\]
This shows that $C_j=\harm{j}$ in \eqref{eq:D2Coeff}
and proves the claim \eqref{eq:HarmCoeff}.

\paragraph{Examples.}

The expression \eqref{eq:BabyHam}
can be used to calculate any
one-loop anomalous dimension within this subsector. 
The generalisation to the non-planar sector is straightforward:
We know that the structure of the non-planar Hamiltonian 
is given by \eqref{eq:Hreduce}. We can therefore act with 
the Hamiltonian density $H_{kl}$ also on non-neighbouring 
sites $k,l$. The non-planar structure is then evaluated according to
\eqref{eq:Hreduce}.

\begin{table}\centering
$\begin{array}[t]{|c|cc|l|}\hline
\Delta_0&L&n& \delta \Delta^P\, [\gym^2 N/\pi^2]\\\hline
4&2&2&\frac{3}{4}^+ \\\hline
5&3&2&\frac{1}{2}^- \\\hline
6&4&2&\frac{1}{8}(5\pm \sqrt{5})^+ \\
 &3&3&\frac{15}{16}^\pm \\
 &2&4&\frac{25}{24}^+ \\\hline
7&5&2&\frac{1}{4}^-, \frac{3}{4}^- \\
 &4&3&\frac{3}{4}^\pm \\
 &3&4&\frac{3}{4}^- \\\hline
\end{array}
\qquad
\begin{array}[t]{|c|cc|l|}\hline
\Delta_0&L&n& \delta \Delta^P\, [\gym^2 N/\pi^2]\\\hline
8&6&2&(64x^3-112x^2+56x-7)^+\\
 &5&3&\frac{1}{32}(25\pm\sqrt{37})^\pm  \\
 &4&4&\frac{23}{24}^\pm, (768x^3-2336x^2+2212x-637)^+ \\
 &3&5&\frac{35}{32}^\pm \\
 &2&6&\frac{49}{40}^+ \\\hline
9&7&2&\frac{1}{4}(2\pm\sqrt{2})^-,\frac{1}{2}^-\\
 &6&3&(256x^3-608x^2+459x-108)^\pm \\
 &5&4&\frac{1}{16}(13\pm\sqrt{41})^-, \frac{1}{96}(97\pm 7\sqrt{5})^\pm \\
 &4&5&\frac{1}{96}(105\pm\sqrt{385})^\pm \\
 &3&6&\frac{11}{12}^-,\frac{227}{160}^\pm \\\hline
\end{array}$
\caption{
The first few states within the $\alSL(2)$ subsector.
The weights of the corresponding primaries are ${[L+n-2;n-2,n-2;0,L-2,0;0,L]}$.
Cubic polynomials indicate three
states with energies given by the roots of the cubic equation.}
\label{tab:BabyEnergy}
\end{table}

We go ahead and find some eigenstates of the Hamiltonian,
see \tabref{tab:BabyEnergy}.
Especially for the `two-body' problems with two sites or 
two excitations (derivatives) one should be able to find exact eigenstates. 
For $L=2$ we know the answer already, it is $\Tr \state{j}$,
where the trace projects to cyclic states, 
effectively removing states with odd $j$. The 
spectrum is given by \eqref{eq:PomeronEnergy}.
Indeed, also for $n=2$ we find exact eigenstates
\[\label{eq:BMNvector}
\Op^L_n=2\cos\frac{\pi n}{L+1}\Tr \,(\osca_1^\dagger)^2\,\state{L}
+\sum_{p=2}^L \cos\frac{\pi n(2p-1)}{L+1}\, \Tr \osca_1^\dagger\osca_p^\dagger\, \state{L},
\]
which are precisely the BMN operators 
with two symmetric-traceless 
vector indices \cite{Beisert:2002tn,Klose:2003tw}.
Their energy is
\[\label{eq:BMNenergy}
E=8\sin^2\frac{\pi n}{L+1},\qquad
\delta D=\frac{\gym^2 N}{\pi^2}\,\sin^2\frac{\pi n}{L+1}.
\]
In analogy to the special, unpaired three-impurity states found in 
\cite{Beisert:2003tq}, we might hope to find 
special states of three sites or with three excitations.
It turns out, that there are no unpaired states 
of three excitations, but there are some for $L=3$.
Empirically we find exactly one unpaired state for $2k$ excitations. 
This state has energy
\[\label{eq:OdderonEnergy}
E=4h(k),\qquad \delta D=\frac{\gym^2 N}{2\pi^2}\,h(k).\]
and weight ${[2k+3;2k,2k;0,3,0;0,3]}$.
The corresponding superconformal primary has weight
\[\label{eq:OdderonReps}
[2k+1;2k-2,2k-2;0,1,0;0,3].\]
Interestingly, the spectrum comprises all harmonic numbers
$h(k)$, whereas the $L=2$ spectrum \eqref{eq:PomeronEnergy}
consists only of the even ones $h(2k)$.

\section{The oscillator picture}
\label{sec:Osc}

In the last section we have made use of 
an oscillator representation for a subsector of operators in
$\superN=4$ SYM.
In this section we will show how to
represent a generic operator of $\superN=4$ SYM 
in terms of excitations of (different) oscillators,
see also \cite{Gunaydin:1985fk}.
We will then describe the action of
the Hamiltonian on these states explicitly.

\subsection{Oscillator representations}
\label{sec:OscRep}

Let us explain the use of oscillators
for fields and generators in terms of the algebra $\alGL(N)$:
We write%
\footnote{Strictly speaking, the oscillators $\osca$ and $\osca^\dagger$ 
are independent. 
Only in one of the real forms of $\alGL(N)$ they would 
be related by conjugation.}
\[\label{eq:OscRep}
J^a{}_b=\osca^\dagger_b\osca^a,
\qquad
\mbox{with }a,b=1,\ldots,N.
\]
Using the commutators
\[\label{eq:algcomm}
\comm{\osca^a}{\osca^\dagger_b}=\delta^a_b,\qquad
\comm{\osca^a}{\osca^b}=\comm{\osca^\dagger_a}{\osca^\dagger_b}=0
\]
it is a straightforward exercise to show that 
$J$ satisfies the $\alGL(N)$ algebra. 

\paragraph{The $\alGL(N)$ invariant vacuum.}

Let us introduce a state $\vac$ defined by
\[\label{eq:OscVac}
\osca^a \vac=0\qquad \mbox{for }a=1,\ldots,N.\]
Then the states
\[\label{eq:OscStates}
\osca^\dagger_{a_1} \ldots \osca^\dagger_{a_k}\vac\]
transform in the 
totally symmetrised product of 
$k$ fundamental representations of $\alSL(N)$ and have $\alGL(1)$ central charge $k$.

\paragraph{A $\alGL(N)$ breaking vacuum.}

Another state $\state{n}$ can be defined by
\[\label{eq:OscVac2}
\osca^a \state{n}=0,
\quad
\osca^\dagger_{a'} \state{n}=0
\qquad \mbox{for }a=1,\ldots,n,\quad a'=n+1,\ldots N.\]
In this case it is more convenient to 
write
\[\label{eq:OscRedef}
\oscb^{\dot a}=\osca^\dagger_{\dot a+n},
\quad
\oscb^\dagger_{\dot a}=-\osca^{\dot a+n},
\quad\mbox{for }\dot a=1,\ldots,n'=N-n\quad\mbox{with }
\comm{\oscb^{\dot{a}}}{\oscb^{\dagger}_{\dot{b}}}=\delta^{\dot a}_{\dot b}
\]
and consider the subalgebra $\alSL(n)\times \alSL(n')$, 
$J^a{}_b=\osca^\dagger_b\osca^a$,
$J^{\dot a}{}_{\dot b}=\oscb^\dagger_{\dot b}\oscb^{\dot a}$,
under which $\state{n}$ is invariant.
Now as the off-diagonal part
$J_{\dot a b}=\osca^\dagger_b\oscb^{\dagger}_{\dot a}$
of the generator $J$ consists of only creation operators,
the state $\state{n}$ transforms in an 
infinite-dimensional representation.
In a generic state
\[\label{eq:OscStates2}
\osca^\dagger_{a_1} \ldots \osca^\dagger_{a_k}
\oscb^{\dagger}_{\dot b_1} \ldots \oscb^{\dagger}_{\dot b_{k'}}
\state{n}\]
the $\alGL(1)$ central charge $J^a{}_a-J^{\dot a}{}_{\dot a}$ is 
$k-k'-n'$ and it labels irreducible modules of 
$\alSL(N)$. The $\alGL(1)$ dilatation charge (dimension) 
$n'J^a{}_a+n J^{\dot a}{}_{\dot a}$
contained in $\alSL(N)$ equals $kn'+k'n+nn'$.
Within each irreducible module there is a state with lowest dimension,
it has either $k=0$ or $k'=0$ depending on the value of the central charge.

\paragraph{Fermionic oscillators.}

Instead of the bosonic commutators
\eqref{eq:algcomm} we might choose 
fermionic oscillators with anticommutators
\[\label{eq:OscFermi}
\acomm{\osca^a}{\osca^\dagger_b}=\delta^a_b,\qquad
\acomm{\osca^a}{\osca^b}=\acomm{\osca^\dagger_a}{\osca^\dagger_b}=0.
\]
In this case it is also straightforward to show that 
the algebra of $\alGL(N)$ is satisfied. 
The oscillator representation $J^a{}_b$ 
splits into $N+1$ 
totally antisymmetric products of the fundamental 
representation of $\alGL(N)$.
This will also be the case for a $\alGL(N)$ breaking vacuum,
though not manifestly.

\subsection{A representation of \texorpdfstring{$\alGL(4|4)$}{gl(4|4)}}
\label{sec:OscRep44}

Keeping all this in mind, let us now have a look at \tabref{tab:SYMfields}.
All representations of $\alSL(2)\times\alSL(2)$
are symmetric tensor products of the fundamental representation, while 
the representations of $\alSL(4)$ are antisymmetric. 
Using two bosonic oscillators 
$(\osca^\alpha,\osca^\dagger_\alpha)$, 
$(\oscb^{\dot \alpha},\oscb^\dagger_{\dot\alpha})$ 
with $\alpha,\dot\alpha=1,2$
and one fermionic oscillator $(\oscc^a,\oscc^\dagger_a)$
with $a=1,2,3,4$
we can thus write \cite{Gunaydin:1985fk}
\<\label{eq:FieldOsc}
\cder^k \fldF \mathrel{}&&\mathrel{\widehat{=}}        
  (\osca^\dagger)^{k+2} 
  (\oscb^\dagger)^{k\phantom{\mathord{}+0}}
  (\oscc^\dagger)^0 
  \vac, \nln
\cder^k \Psi \mathrel{}&&\mathrel{\widehat{=}}     
  (\osca^\dagger)^{k+1} 
  (\oscb^\dagger)^{k\phantom{\mathord{}+0}}
  (\oscc^\dagger)^1 
  \vac, \nln
\cder^k \Phi \mathrel{}&&\mathrel{\widehat{=}}     
  (\osca^\dagger)^{k\phantom{\mathord{}+0}} 
  (\oscb^\dagger)^{k\phantom{\mathord{}+0}} 
  (\oscc^\dagger)^2 
  \vac, \nln
\cder^k \bar\Psi \mathrel{}&&\mathrel{\widehat{=}} 
  (\osca^\dagger)^{k\phantom{\mathord{}+0}} 
  (\oscb^\dagger)^{k+1} 
  (\oscc^\dagger)^3 
  \vac, \nln
\cder^k \bar \fldF \mathrel{}&&\mathrel{\widehat{=}}   
  (\osca^\dagger)^{k\phantom{\mathord{}+0}}
  (\oscb^\dagger)^{k+2} 
  (\oscc^\dagger)^4 
  \vac .
\>
Each of the oscillators $\osca^\dagger_\alpha,\oscb^\dagger_{\dot\alpha},\oscc^\dagger_a$
carries one of the $\alSL(2)^2,\alSL(4)$ spinor indices of the fields,
for example
\[\label{eq:FieldOscEx}
\cder_{\alpha\dot\beta}\,\cder_{\gamma\dot\delta}\,\Phi_{ab}=
\osca^\dagger_\alpha\osca^\dagger_\gamma
\oscb^\dagger_{\dot\beta}\oscb^\dagger_{\dot\delta}
\oscc^\dagger_a\oscc^\dagger_b\vac.
\]
The statistics of the oscillators automatically symmetrises the
indices in the desired way.
The non-vanishing commutators of oscillators are taken to be
\<\label{eq:OscComm}
\comm{\osca^{\alpha}}{\osca^\dagger_{\beta}}\eq \delta^\alpha_\beta,
\nln
\comm{\oscb^{\dot\alpha}}{\oscb^\dagger_{\dot\beta}}\eq \delta^{\dot\alpha}_{\dot\beta},
\nln
\acomm{\oscc^{a}}{\oscc^\dagger_{b}}\eq\delta^a_b.
\>
The canonical forms for the generators of the two $\alSL(2)$
and $\alSL(4)$ are
\<\label{eq:GL44a}
L^{\alpha}{}_{\beta}\eq\osca^\dagger_{\beta}\osca^{\alpha}
-\half \delta^\alpha_\beta\osca^\dagger_{\gamma}\osca^{\gamma},
\nln
\dot L^{\dot\alpha}{}_{\dot\beta}\eq\oscb^\dagger_{\dot\beta}\oscb^{\dot\alpha}
-\half \delta^{\dot\alpha}_{\dot\beta}\oscb^\dagger_{\dot\gamma}\oscb^{\dot\gamma},
\nln
R^{a}{}_{b}\eq\oscc^\dagger_{b}\oscc^{a}
-\quarter \delta^{a}_{b}\oscc^\dagger_{c}\oscc^{c}.
\>
Under these the fields \eqref{eq:FieldOsc} transform canonically.
We write the corresponding three $\alGL(1)$ charges as
\<\label{eq:GL44b}
D\eq 
1+\half \osca^\dagger_{\gamma}\osca^{\gamma}
+\half \oscb^\dagger_{\dot\gamma}\oscb^{\dot\gamma},
\nln
C\eq 
1-\half \osca^\dagger_{\gamma}\osca^{\gamma}
+\half \oscb^\dagger_{\dot\gamma}\oscb^{\dot\gamma}
-\half \oscc^\dagger_{c}\oscc^{c},
\nln
B\eq \phantom{1+\mathord{}}\half \osca^\dagger_{\gamma}\osca^{\gamma}
-\half \oscb^\dagger_{\dot\gamma}\oscb^{\dot\gamma}.
\>
Assuming that the oscillators 
$(\osca,\oscb^\dagger,\oscc)$ and 
$(\osca^\dagger,\oscb,\oscc^\dagger)$
transform in fundamental and 
conjugate fundamental representations
of $\alGL(4|4)$ we write down the remaining 
off-diagonal generators 
according to \eqref{eq:OscRep}
\[\label{eq:GL44c}
\arraycolsep0pt
\begin{array}{rclcrcl}
Q^a{}_{\alpha}\eq\osca^\dagger_\alpha \oscc^{a},&\qquad&
S^{\alpha}{}_a\eq\oscc^\dagger_{a} \osca^\alpha , 
\\[3pt]
\dot Q_{\dot\alpha a}\eq \oscb^\dagger_{\dot\alpha} \oscc^\dagger_{a} ,&&
\dot S^{\dot\alpha a}\eq \oscb^{\dot\alpha} \oscc^{a} ,
\\[3pt]
P_{\alpha \dot \beta}\eq\osca^\dagger_{\alpha}\oscb^\dagger_{\dot \beta} ,&&
K^{\alpha \dot \beta}\eq\osca^{\alpha}\oscb^{\dot \beta}.
\end{array}
\]
Quite naturally the algebra $\alGL(4|4)$ 
is realised by the generators
\eqref{eq:GL44a},\eqref{eq:GL44b}%
\footnote{
Note that a shift of $B$ by a constant ($-1$) does not modify the algebra.
Then the $1$ in $D,C,B$ can be absorbed into 
$1+\half \oscb^\dagger_{\dot\gamma}\oscb^{\dot\gamma}=\half \oscb^{\dot\gamma}\oscb^\dagger_{\dot\gamma}$
to yield a canonical form.},\eqref{eq:GL44c}.
We have written this in a $\alSL(2)^2\times\alSL(4)$ 
covariant way. In fact one combine the indices $a$ and $\alpha$ into
a superindex and obtain a manifest $\alSL(2)\times\alSL(2|4)$
notation. The generators with two lower or two upper indices,
$P,\dot Q,K,\dot S$, 
together with the remaining charges complete the $\alGL(4|4)$ algebra.

First, we note that all fields
\eqref{eq:FieldOsc} are uncharged with respect to
the central charge $C$, it can therefore be 
dropped leading to $\alSL(4|4)$.
Furthermore, in $\alSL(4|4)$ the generator $B$ never appears 
in commutators and can be projected out,
this algebra is $\alPSL(4|4)$. 
The generators \eqref{eq:GL44a},\eqref{eq:GL44b},\eqref{eq:GL44c}
form a specific irreducible representation of $\alPSL(4|4)$. As they 
transform the fields \eqref{eq:FieldOsc} of $\superN=4$ SYM among 
themselves, this representation 
has primary weight \eqref{eq:Vfund}
\[\label{eq:Vfund2}
w\indup{F}=[1;0,0;0,1,0;0,1].\]
As an aside, in this representation it can be explicitly shown that the value of 
the quadratic Casimir operator \eqref{eq:gl44Casimir} is zero
\[\label{eq:CasimirVF}
J^2 V\indup{F}=0.
\]
In the same manner it can be shown that 
the value of the quadratic Casimir in the module $V_j$,
\eqref{eq:CasimirVj}, is $j(j+1)$.

\paragraph{Physical vacuum.}

As in the case of the conformal subalgebra $\alSL(4)$ we can split the 
flavour $\alSL(4)$ into $\alSL(2)\times\alSL(2)$.
In that case we define
\[\label{eq:SymSplit}
\oscd^\dagger_1=\oscc^3,\quad \oscd^\dagger_2=\oscc^4,\qquad
\oscd^1=\oscc^\dagger_3,\quad \oscd^2=\oscc^\dagger_4,
\]
and use a vacuum $\state{Z}$ annihilated by $\osca_{1,2},\oscb_{1,2},\oscc_{1,2},\oscd_{1,2}$.
It is related to $\vac$ by 
\[\label{eq:SymSplitVac}
\state{Z}=\oscc^\dagger_3\oscc^\dagger_4\vac.
\]
The benefit of this vacuum is that it is not charged under
the central charge $C$ and thus physical. 
It corresponds to the primary weight $[1;0,0;0,1,0;0,1]$ of 
the module $V\indup{F}$.
Furthermore it is the vacuum used in the BMN correspondence
\cite{Berenstein:2002jq}
and one of the Bethe ans\"atze in \cite{Beisert:2003yb}.
The drawback is that it is not 
invariant under the full $\alSL(4)$ flavour algebra, but only
under a subgroup $\alSL(2)\times\alSL(2)$. 
The expressions for the $\alGL(4|4)$ generators
thus complicate. However, if the indices 
$(a,\alpha)$ and $(\dot a,\dot \alpha)$ are combined in two
superindices, we have a manifest 
$\alPSL(2|2)\times\alPSL(2|2)$ covariance.

\paragraph{Weights and excitations.}

In this context it is useful to know how
to represent an operator 
with a given weight 
\[\label{eq:OscWeight}
w=[\Delta_0;s_1,s_2;q_1,p,q_2;B,L]\]
in terms of excitations of the oscillators.
We introduce a vacuum operator $\state{0,L}$ which is the tensor product
of $L$ vacua $\vac$.
The oscillators 
$\osca^\dagger_{s,\alpha},\oscb^\dagger_{s,\dot\alpha},\oscc^\dagger_{s,a}$.
now act on site $s$, where
commutators of two oscillators 
vanish unless the sites agree.
A generic state is written as
\[
(\osca^\dagger)^{n_{\osca}}(\oscb^\dagger)^{n_{\oscb}}(\oscc^\dagger)^{n_{\oscc}}\state{0,L}
\]
By considering the weights of the oscillators
as well as the central charge constraint, we find 
the number of excitations 
\[\label{eq:OscExciteab}
n_{\osca}=\left(\begin{array}{c}
\half \Delta_0+\half B-\half L+\half s_1\\
\half \Delta_0+\half B-\half L-\half s_1
\end{array}\right),\quad
n_{\oscb}=\left(\begin{array}{c}
\half \Delta_0-\half B-\half L+\half s_2\\
\half \Delta_0-\half B-\half L-\half s_2
\end{array}\right).
\]
and
\[\label{eq:OscExcitec}
n_{\oscc}=\left(\begin{array}{c}
\half L-\half B-\half p-\sfrac{3}{4} q_1-\sfrac{1}{4}q_2\\
\half L-\half B-\half p+\sfrac{1}{4} q_1-\sfrac{1}{4}q_2\\
\half L-\half B+\half p+\sfrac{1}{4} q_1-\sfrac{1}{4}q_2\\
\half L-\half B+\half p+\sfrac{1}{4} q_1+\sfrac{3}{4}q_2
\end{array}\right).
\]
If the physical vacuum $\state{Z}$ is used instead of the 
$\alSL(4)$ invariant vacuum, the numbers of excitations
of the $\alSL(4)$ oscillators are given by
\[\label{eq:OscExcitecd}
n_{\oscc}=\left(\begin{array}{c}
\half L-\half B-\half p-\sfrac{3}{4} q_1-\sfrac{1}{4}q_2\\
\half L-\half B-\half p+\sfrac{1}{4} q_1-\sfrac{1}{4}q_2
\end{array}\right),\quad
n_{\oscd}=\left(\begin{array}{c}
\half L+\half B-\half p-\sfrac{1}{4}q_1+\sfrac{1}{4}q_2\\
\half L+\half B-\half p-\sfrac{1}{4}q_1-\sfrac{3}{4}q_2
\end{array}\right).
\]

\subsection{The harmonic action}
\label{sec:OscHamil}

The $\alPSL(4|4)$ invariant Hamiltonian density $H_{12}$ is given 
by some function of $J_{12}$,
see \eqref{eq:FunctionOfJ}
\[\label{eq:HarmHam}
H_{12}=2\harm{J_{12}}.
\]
We will now describe explicitly how $H_{12}$ acts
on a state of two-sites.

\paragraph{Invariant action.}

We will investigate the action 
of a generic function $f(J_{12})$ on two oscillator sites.
As the dimension of any explicit state is a finite number,
we can express $f(J_{12})$ as a polynomial in the quadratic
Casimir, \eqref{eq:FuncOfJ}.
Let us introduce a collective oscillator 
$\oscA^\dagger_{A}=(\osca^\dagger_{\alpha},\oscb^\dagger_{\dot\alpha},\oscc^\dagger_{a})$.
A general state in $V\indup{F}\times V\indup{F}$ can be written as
\[\label{eq:OscChainState}
\state{s_1,\ldots,s_n;A}=
\oscA^\dagger_{s_1,A_1}
\ldots
\oscA^\dagger_{s_{n},A_{n}}
\state{00}
\]
subject to the central charge constraints $C_1\state{X}=C_2\state{X}=0$.
The label $s_k=1,2$ determines the site on which the $k$-th oscillator acts.
It is easily seen that the Casimir operator $J_{12}^2$ conserves the number of
each type of oscillator; it can however
move oscillators between both sites.
Therefore the action of $f(J_{12})$ is 
\[\label{eq:FuncOnState}
f(J_{12})\state{s_1,\ldots,s_n;A}=
\sum_{s'_1,\ldots s'_{n}}
c_{s,s',A}\,
\delta_{C_1,0}\delta_{C_2,0}\,
\state{s'_1,\ldots,s'_n;A}
\]
with some coefficients $c_{s,s',A}$.
The sums go over the sites $1,2$
and $\delta_{C_1,0}$, $\delta_{C_2,0}$ project 
to states where the central charge at each site is zero.
In view of the fact that oscillators represent
indices of fields, see \eqref{eq:FieldOscEx}, a generic 
invariant operators $f(J_{12})$ acts
on two fields by moving indices between them.

\paragraph{Harmonic action.}

The action of the harmonic numbers
within the Hamiltonian density $H_{12}=2\harm{J_{12}}$
turns out to be particularly simple.
It does not depend on the 
types of oscillators $A_k$,
but only on the number of oscillators which change the site
\[\label{eq:HamAction}
H_{12}\state{s_1,\ldots,s_n;A}=
\sum_{s'_1,\ldots s'_{n}}
c_{n,n_{12},n_{21}}\,
\delta_{C_1,0}\delta_{C_2,0}\,
\state{s'_1,\ldots,s'_n;A}.
\]
Here $n_{12},n_{21}$ count the number 
of oscillators hopping 
from site $1$ to $2$ or vice versa.
The coefficients $c_{n,n_{12},n_{21}}$ are given by
\[\label{eq:HamCoeff}
c_{n,n_{12},n_{21}}=
 (-1)^{1+n_{12}n_{21}}
\frac{\Gamma\bigbrk{\half (n_{12}+n_{21})}\Gamma\bigbrk{1+\half (n-n_{12}-n_{21})}}{\Gamma\bigbrk{1+\half n}}.
\]
In the special case of no oscillator hopping we find
\[\label{eq:HamCoeff00}
c_{n,0,0}= \harm{\half n},\]
which can be regarded as a regularisation of \eqref{eq:HamCoeff}.
We will refer to this action given by 
\eqref{eq:HamAction},\eqref{eq:HamCoeff},\eqref{eq:HamCoeff00}
as the `harmonic action'.

\paragraph{Proof.}

To prove that $H_{12}$ is given by 
\eqref{eq:HamAction},\eqref{eq:HamCoeff},\eqref{eq:HamCoeff00}
it suffices to show
\[\label{eq:ProveThis}
\comm{J_{12}}{H_{12}}=0,\qquad
H_{12}V_{j}=2\harm{j} V_j.
\]
The invariance of $H_{12}$ under 
the subalgebra $\alSL(2)\times\alSL(2|4)$ 
given by the generators $\dot L,L,R,D,Q,S$
is straightforward:
These generators only change the types of oscillators,
whereas the harmonic action does not depend on that.
In contrast, the remaining generators
$K,P,\dot Q,\dot S$ change the number of oscillators by two.

Let us act with $P_{12,\alpha\dot\beta}$ on a generic state
\[\label{eq:PonState}
P_{12,\alpha\dot\beta}
\state{s_1,\ldots,s_n;A}=
\state{1,1,s_1,\ldots,s_n;A'}+
\state{2,2,s_1,\ldots,s_n;A'},
\]
and get a state with two new oscillators, $A'=(\alpha,\dot\beta,A)$.
The action of the Hamiltonian density \eqref{eq:HamAction}
yields eight terms.
In two of these terms both new oscillators end up at site $1$
\[\label{eq:PCommGood}
c_{n+2,n_{12},n_{21}}\state{1,1,s'_1,\ldots,s'_n;A'}+
c_{n+2,n_{12},n_{21}+2}\state{1,1,s'_1,\ldots,s'_n;A'}.
\]
Here $n_{12},n_{21}$ refer only to the hopping of the old oscillators.
Eq. \eqref{eq:HamCoeff} 
can be used to combine the two coefficients in one
\[\label{eq:HarmCoeffAdd}
c_{n+2,n_{12},n_{21}}+c_{n+2,n_{12},n_{21}+2}=c_{n,n_{12},n_{21}}.
\]
We pull the additional two oscillators out of the state and get
\[\label{eq:PCommDone}
\bigbrk{c_{n+2,n_{12},n_{21}}+c_{n+2,n_{12},n_{21}+2}}\state{1,1,s'_1,\ldots,s'_n;A'}=
P_{1,\alpha\dot\beta}\, c_{n,n_{12},n_{21}}\state{s'_1,\ldots,s'_n;A}
\]
Summing up all contributions therefore yields
$P_{1,\alpha\dot\beta}H_{12}\state{s_1,\ldots,s_n;A}$. 
If both new oscillators end up at site $2$ we get an equivalent result.
It remains to be shown that the other four terms cancel.
Two of these are
\[\label{eq:PCommBad}
c_{n+2,n_{12},n_{21}+1}\state{1,2,s'_1,\ldots,s'_n;A'}+
c_{n+2,n_{12}+1,n_{21}}\state{1,2,s'_1,\ldots,s'_n;A'}.
\]
We see that the absolute values in \eqref{eq:HamCoeff}
match for $c_{n+2,n_{12},n_{21}+1}$ and $c_{n+2,n_{12}+1,n_{21}}$,
we sum up the signs
\[\label{eq:PCommCancel}
(-1)^{1+n_{12}n_{21}+n_{12}}+(-1)^{1+n_{12}n_{21}+n_{21}}=
(-1)^{1+n_{12}n_{21}}\bigbrk{(-1)^{n_{12}}+(-1)^{n_{21}}}.
\]
Now, oscillators always hop in pairs due to the central charge constraint.
One of the new oscillators has changed the site, so 
the number of old oscillators changing site must be odd. 
The above two signs must be opposite and cancel in the sum.
The same is true for the remaining two terms.
This concludes the proof for $\comm{P_{12,\alpha\dot\beta}}{H_{12}}=0$
and similarly, for invariance under $\dot Q$.
To prove invariance under $K,\dot S$ we note that
these generators remove two oscillators from one of the two sites.
Assume it will remove the first two oscillators from a state
(for each two oscillators that are removed, the argument will be the same).
Now, the argument is essentially the
same as the proof for $P_{12,\alpha\dot\beta}$ 
read in the opposite direction.

To prove that the eigenvalues of $H_{12}$ are given by $2\harm{j}$
we act on a special state within $V_j$.
We define
\[\label{eq:SpinStatePart}
\state{j+2,k}=
\frac{
(\osca^\dagger_{1,1})^{k+2}(\oscb^\dagger_{1,1})^k
(\osca^\dagger_{2,1})^{j-k+2}(\oscb^\dagger_{2,1})^{j-k}}{(k+2)!k!(j-k+2)!(j-k)!}
\,\state{00}
\]
which corresponds to $\cder^k \fldF \,\cder^{j-k} \fldF$.
Then 
\[\label{eq:SomeSpinState}
\state{j+2}=
\sum_{k=0}^j (-1)^k\,
\state{j+2,k}
\]
is a representative of $V_{j+2}$.
It is therefore an eigenstate of $H_{12}$ and
we can choose to calculate only the
coefficient of $\state{j+2,0}$ in $H_{12}\state{j+2}$.
Using some combinatorics we find the coefficient
\[\label{eq:SomeSpinCoeff}
h(j+2)+
\sum_{l=0}^2
\sum_{k=0}^{j}
(-1)^{1+k+l}\,
\frac{\delta_{k+l\neq 0}}{k+l}\,
\frac{2}{l!(2-l)!}\,
\frac{j!}{k!(j-k)!}
=2h(j+2),
\]
where $l$ represents the number of 
oscillators $\osca^\dagger_{1,2}$ hopping to site $1$.
For $j=0,1$ we define the states
\<\label{eq:SomeLowSpinState}
\state{j=0}\eq a^\dagger_{1,1}a^\dagger_{1,1}a^\dagger_{2,1}a^\dagger_{2,1}\state{00}
+a^\dagger_{2,1}a^\dagger_{2,1}a^\dagger_{1,1}a^\dagger_{1,1}\state{00}
-2a^\dagger_{1,1}a^\dagger_{2,1}a^\dagger_{1,1}a^\dagger_{2,1}\state{00},
\nln
\state{j=1}\eq a^\dagger_{1,1}a^\dagger_{1,1}a^\dagger_{2,1}a^\dagger_{2,1}\state{00}
-a^\dagger_{2,1}a^\dagger_{2,1}a^\dagger_{1,1}a^\dagger_{1,1}\state{00}.
\>
It is a straightforward exercise to show that $H_{12}\state{j=0}=0$
and $H_{12}\state{j=1}=2\state{j=1}$.
This concludes the proof of \eqref{eq:ProveThis}.

\paragraph{Physical vacuum.}

We can also describe the harmonic action using the
primary vacuum $\state{Z}$. 
Then the collective oscillator is 
$\oscA^\dagger_{A}=(\osca^\dagger_{\alpha},\oscb^\dagger_{\dot\alpha},\oscc^\dagger_{a},\oscd^\dagger_{\dot a})$.
A generic state in $V\indup{F}\times V\indup{F}$ is 
defined in analogy to \eqref{eq:OscChainState}.
Interestingly, we find that
the action of the Hamiltonian density 
is obtained using exactly the same expressions
\eqref{eq:HamAction},\eqref{eq:HamCoeff},\eqref{eq:HamCoeff00}.
Invariance of $H_{12}$ can be shown as above
and the proof for the eigenvalues is very similar.
We might use the states
\[\label{eq:OtherSpinStatePart}
\state{j,k}=
\frac{(\osca^\dagger_{1,1})^{k}(\oscb^\dagger_{1,1})^k
(\osca^\dagger_{2,1})^{j-k}(\oscb^\dagger_{2,1})^{j-k}}{{k!^2(j-k)!^2}}
\,\state{ZZ},
\]
which correspond to the state $k!(j-k)! (\osca^\dagger_1)^k (\osca^\dagger_2)^{j-k}\state{00}$ 
of \secref{sec:Baby}.
Then 
\[\label{eq:OtherSpinState}
\state{j}=\sum_{k=0}^j (-1)^k\,\state{j,k}
\]
corresponds to $j!\,\state{j}$ 
of \eqref{eq:BabyHWState} and belongs to the module $V_j$. 
As above we consider
only the coefficient of $\state{j,0}$ in $H_{12}\state{j}$.
It is given by 
\[\label{eq:OtherSpinCoeff}
h(j)+
\sum_{k=1}^{j}
\frac{(-1)^{1+k} j!}{k\,k!(j-k)!}
=2h(j),
\]
which proves that $H_{12}=2h(J_{12})$.

\subsection{Some examples}
\label{sec:OscEx}

We will now determine the planar anomalous dimension of a couple of operators
to demonstrate how to apply the above Hamiltonian.

\paragraph{Konishi.}
The Konishi operator has weight $[2;0,0;0,0,0;0,2]$. Using 
\eqref{eq:OscExciteab},\eqref{eq:OscExcitec}
we find that we have to excite each of the four oscillators $\oscc$ once. 
There must be two oscillators on each site due to the central charge constraint
and the three distinct configurations are
\[\label{eq:XXbStates}
\arraycolsep0pt\begin{array}{rcl}
\state{2112}\eq 
\oscc^\dagger_{2,1}\oscc^\dagger_{1,2}
\oscc^\dagger_{1,3}\oscc^\dagger_{2,4}
\state{00}= \Tr \bar XX
,\\[3pt]
\state{1212}\eq
\oscc^\dagger_{1,1}\oscc^\dagger_{2,2}
\oscc^\dagger_{1,3}\oscc^\dagger_{2,4}
\state{00}= \Tr \bar YY
,\\[3pt]
\state{1122}\eq
\oscc^\dagger_{1,1}\oscc^\dagger_{1,2}
\oscc^\dagger_{2,3}\oscc^\dagger_{2,4}
\state{00}= \Tr \bar ZZ.
\end{array}
\]
These can also be written in terms of 
three complex scalars $X,Y,Z$ of $\superN=4$ SYM and their conjugates.
We can define them as
\[\label{eq:XYZRep}
\arraycolsep0pt
\begin{array}{rclcrclcrcl}
X\eq\oscc^\dagger_1 \oscc^\dagger_4 \vac,&\quad&
Y\eq\oscc^\dagger_2 \oscc^\dagger_4 \vac,&\quad&
Z\eq\oscc^\dagger_3 \oscc^\dagger_4 \vac,\\[3pt]
\bar X\eq\oscc^\dagger_2 \oscc^\dagger_3 \vac,&&
\bar Y\eq\oscc^\dagger_3 \oscc^\dagger_1 \vac,&&
\bar Z\eq\oscc^\dagger_1 \oscc^\dagger_2 \vac.
\end{array}
\]
Let us now act with $H_{12}$ on these states, we find
\<\label{eq:HOnXX}
H_{12}\state{1122}\eq
c_{4,0,0}\state{1122}
+c_{4,1,1}\state{1221}
+c_{4,1,1}\state{1212}
\nl
+c_{4,1,1}\state{2121}
+c_{4,1,1}\state{2112}
+c_{4,2,2}\state{2211}
\nle
\sfrac{3}{2}\state{1122}
+\sfrac{1}{2}\state{1221}
+\sfrac{1}{2}\state{1212}
+\sfrac{1}{2}\state{2121}
+\sfrac{1}{2}\state{2112}
-\sfrac{1}{2}\state{2211}
\nle
\state{2112}+\state{1212}+\state{1122}
\>
using \eqref{eq:HamAction},\eqref{eq:HamCoeff},\eqref{eq:HamCoeff00}
and cyclicity of the trace.
Evaluating the Hamiltonian for the
remaining two states $\state{1212}$ and $\state{1221}$
we find the energy matrix
\[\label{eq:HOnStateMatrix}
H=\left(\begin{array}{ccc}2&2&2\\2&2&2\\2&2&2\end{array}\right),
\]
the factor of $2$ is due to $H=H_{12}+H_{21}$.
One eigenstate is
\[\label{eq:XXKonishi}
\mathcal{K}=\state{2112}+\state{1212}+\state{1122}
=\Tr \bar X X+\Tr \bar YY+\Tr \bar ZZ.
\]
with energy
\[\label{eq:KonEng}
E=6,\qquad
\delta\Delta=E\times \frac{\gym^2N}{8\pi^2}=\frac{3\gym^2N}{4\pi^2}
\]
which clearly corresponds to the Konishi operator.
The other two, $\state{2112}-\state{1122}$ and $\state{1212}-\state{1122}$,
have vanishing energy and correspond to half-BPS operators.

\paragraph{Physical vacuum.}

Let us repeat this example using the physical vacuum $\state{Z}$.
Here, we can define
\[\label{eq:XYZRep2}
\arraycolsep0pt
\begin{array}{rclcrclcrcl}
X\eq \oscc^\dagger_1 \oscd^\dagger_1\state{Z},&\quad&
Y\eq \oscc^\dagger_2 \oscd^\dagger_1 \state{Z},&\quad&
Z\eq\state{Z},\\[3pt]
\bar X\eq \oscd^\dagger_2 \oscc^\dagger_2 \state{Z},&&
\bar Y\eq \oscc^\dagger_1 \oscd^\dagger_2 \state{Z},&&
\bar Z\eq \oscc^\dagger_1 \oscc^\dagger_2\oscd^\dagger_2 \oscd^\dagger_1 \state{Z}.
\end{array}
\]
The three states are
\[\label{eq:XXbStates2}
\arraycolsep0pt\begin{array}{rcl}
\state{1212'}\eq
\oscc^\dagger_{1,1}\oscc^\dagger_{2,2}
\oscd^\dagger_{1,1}\oscd^\dagger_{2,2}
\state{ZZ}=\Tr \bar XX,
\\[3pt]
\state{1221'}\eq
\oscc^\dagger_{1,1}\oscc^\dagger_{2,2}
\oscd^\dagger_{2,1}\oscd^\dagger_{1,2}
\state{ZZ}=\Tr \bar YY,
\\[3pt]
\state{1111'}\eq \oscc^\dagger_{1,1}\oscc^\dagger_{1,2}
\oscd^\dagger_{1,1}\oscd^\dagger_{1,2}
\state{ZZ}=-\Tr \bar ZZ.
\end{array}
\]
In the same way as above we get the energy matrix 
\[\label{eq:HOnStateMatrix2}
H=\left(\begin{array}{rrr}2&2&-2\\2&2&-2\\-2&-2&2\end{array}\right).
\]
The eigenstates and energies are the same as above, e.g.
\[\label{eq:XXKonishi2}
\mathcal{K}=\state{1212'}+\state{1221'}-\state{1111'}
=\Tr \bar XX+\Tr \bar YY+\Tr \bar ZZ.\]
%

\paragraph{Pseudovacua.}

In the previous paragraph we have worked with the physical vacuum
made from the scalar field $\state{Z}$.
This vacuum is the ground state, 
it is a protected half-BPS state with zero energy.
There are, however, similar configurations where 
we assume the same field at each site. 
We find three possible fields to choose (including the above)
\[
\state{Z}=\oscc^\dagger_3\oscc^\dagger_4\vac,
\qquad
\state{\Psi}=\osca^\dagger_1\oscc^\dagger_4\vac,
\qquad
\state{\fldF}=\osca^\dagger_1\osca^\dagger_1\vac.
\]
In each case the states of two sites
$\state{ZZ},\state{\Psi\Psi},\state{\fldF\fldF}$ 
are eigenstates of the Hamiltonian density,
the corresponding eigenvalues are $0,2,3$.
Thus the energies of the vacua 
$\state{Z,L},\state{\Psi,L},\state{\fldF,L}$
constructed from $L$ such fields are
\[
E_{Z}=0,\qquad
E_{\Psi}=2L,\qquad
E_{\fldF}=3L,
\]
where $\state{\Psi,L}$ exists only for odd $L$.
The $\alSL(4|4)$ weights of these vacua 
and the corresponding primaries are given by
\[\arraycolsep0pt\begin{array}{rclcrcl}
w_Z\eq[L;0,0;0,L,0;0,L],&\quad&
w_{Z,0}\eq[L;0,0;0,L,0;0,L],
\\[3pt]
w_\Psi\eq[\sfrac{3}{2}L;L,0;0,0,L;\half L,L],&\quad&
w_{\Psi,0}\eq[\sfrac{3}{2}L-\frac{5}{2};L-3,0;0,0,L-3;\half L-\sfrac{3}{2},L-1],
\\[3pt]
w_{\fldF}\eq[2L;2L,0;0,0,0;L,L],&\quad&
w_{\fldF,0}\eq[2L-2;2L-4,0;0,0,0;L-2,L].
\end{array}
\]
Of these three states, only $\state{Z,L}$ is stable,
it does not receive corrections at higher loops.
The other two states are expected to receive corrections to their form;
to determine their higher-loop energy becomes a non-trivial issue.

\paragraph{Spectrum of operators with a low dimension.}

We have used the harmonic action to compute the planar one-loop spectrum
of low-lying states in $\superN=4$ SYM. 
First of all we have determined the primary states using 
the algorithm proposed in \cite{Bianchi:2003wx}.
In analogy to the sieve of Eratostene
the algorithm subsequently removes descendants from the
set of all states. What remains, are the primary states.
For the primaries we have determined the
number of oscillator excitations using 
\eqref{eq:OscExciteab},\eqref{eq:OscExcitec},\eqref{eq:OscExcitecd}.
Next we have spread the oscillators on the sites in all 
possible distinct ways.
The harmonic action 
\eqref{eq:HamAction},\eqref{eq:HamCoeff},\eqref{eq:HamCoeff00}
then yields an energy matrix that was 
subsequently diagonalised. 
For all the descendants that were 
removed in the sieve algorithm, we remove the corresponding energy eigenvalue.
The remaining eigenvalue is the one-loop planar anomalous dimension 
of the primary operator.

\begin{table}\centering
$\begin{array}{|c|cccc|l|}\hline
\Delta_0&\alSL(2)^2&\alSL(4)&B&L&\delta\Delta^P\, [\gym^2 N/\pi^2]\\\hline
2&[0,0]&[0,2,0]&0&2&0^+ \\
 &[0,0]&[0,0,0]&0&2&\frac{3}{4}^+ \\
\hline
3&[0,0]&[0,3,0]&0&3&0^- \\
 &[0,0]&[0,1,0]&0&3&\frac{1}{2}^- \\
\hline
4&[0,0]&[0,4,0]&0&4&0^+ \\
 &[0,0]&[0,2,0]&0&4&\frac{1}{8}(5\pm \sqrt{5})^+ \\
 &[0,0]&[1,0,1]&0&4&\frac{3}{4}^- \\
 &[0,0]&[0,0,0]&0&4&\frac{1}{16}(13\pm \sqrt{41})^+ \\
 &[2,0]&[0,0,0]&1&3&\frac{9}{8}^- \hfill+\mathord{\mbox{conj.}}\\
 &[1,1]&[0,1,0]&0&3&\frac{15}{16}^{\pm}\\
 &[2,2]&[0,0,0]&0&2&\frac{25}{24}^{+}\\
\hline
5&[0,0]&[0,5,0]&0&5&0^- \\
 &[0,0]&[0,3,0]&0&5&\frac{1}{4}^-,\frac{3}{4}^- \\
 &[0,0]&[1,1,1]&0&5&\frac{5}{8}^\pm \\
 &[0,0]&[0,0,2]&0&5&\frac{1}{8}(7\pm \sqrt{13})^+  \hfill+\mathord{\mbox{conj.}} \\
 &[0,0]&[0,1,0]&0&5&\frac{5}{4}^-,\frac{5}{4}^-,\frac{1}{8}(5\pm \sqrt{5})^- \\
 &[2,0]&[0,0,2]&1&4&\frac{5}{4}^- \hfill+\mathord{\mbox{conj.}}\\
 &[2,0]&[0,1,0]&1&4&\frac{1}{8}(8\pm \sqrt{2})^+  \hfill+\mathord{\mbox{conj.}} \\
 &[1,1]&[0,2,0]&0&4&\frac{3}{4}^\pm \\
 &[1,1]&[1,0,1]&0&4&\frac{5}{8}^\pm,\frac{5}{4}^\pm \\
 &[1,1]&[0,0,0]&0&4&\frac{9}{8}^\pm \\
 &[2,2]&[0,1,0]&0&3&\frac{3}{4}^-\\
\hline
5.5&[1,0]&[0,2,1]&\half&5&1^\pm  \hfill+\mathord{\mbox{conj.}}\\
   &[1,0]&[1,1,0]&\half&5&\frac{1}{8}(8\pm\sqrt{2})^\pm  \hfill+\mathord{\mbox{conj.}}\\  
   &[1,0]&[0,0,1]&\half&5&\frac{1}{32}(35\pm\sqrt{5})^\pm  \hfill+\mathord{\mbox{conj.}}\\
   &[2,1]&[0,1,1]&\half&4&\frac{9}{8}^\pm  \hfill+\mathord{\mbox{conj.}}\\
   &[2,1]&[1,0,0]&\half&4&\frac{1}{32}(37\pm\sqrt{37})^\pm  \hfill+\mathord{\mbox{conj.}}\\
   &[3,2]&[0,0,1]&\half&3&\frac{5}{4}^\pm  \hfill+\mathord{\mbox{conj.}}\\
\hline
\end{array}$
\caption{All one-loop planar anomalous dimensions of primary operators with $\Delta_0\leq 5.5$.
Positive parity $P$ indicates a state that survives
the projection to gauge groups $\grSO(N)$, $\grSp(N)$,
parity $\pm$ indicates a degenerate pair.
The label `+conj.' indicates conjugate states with
$\alSL(2)^2,\alSL(4)$ labels reversed and opposite chirality $B$.}
\label{tab:anotab}
\end{table}
The single-trace superconformal primaries of $\superN=4$ SYM 
for $\Delta_0\leq 5.5$ and their planar one-loop anomalous dimensions 
are given in \tabref{tab:anotab}.%
\footnote{The table was computed as follows. 
A \texttt{C++} programme was used to determine all
superconformal primary operators up to and including 
classical dimension $5.5$ as well as their descendants. 
For the details of the implementation 
of the sieve algorithm see \cite{Beisert:2003te}.
In a \texttt{Mathematica} programme all operators with 
a given number of excitations of the 
oscillators were collected; this involved up to hundreds of 
states for $\Delta_0=5.5$. In a second step, 
the harmonic action was applied to all these operators to
determine the matrix of anomalous dimensions;
this took a few minutes. 
The relevant eigenvalue corresponding to the primary 
operator was sorted out by hand.}
For a given primary weight we write the anomalous dimensions 
along with a parity $P$.
Here, parity is defined such that 
for a $\grSO(N)$ or $\grSp(N)$ gauge group 
the states with negative parity are projected out.%
\footnote{Therefore, our definition of parity differs 
from the one of \cite{Beisert:2003tq} by a factor of $(-1)^L$.} 
Parity $P=\pm$ indicates a pair of states with opposite parity
but degenerate energy.
Furthermore, we have indicated states with conjugate representations
for which the order of $\alSL(2)^2$ and $\alSL(4)$ labels
as well as the chirality $B$ is inverted.

There are two important points to note looking at the 
energies in \tabref{tab:anotab}.
Firstly, we note the appearance of
paired states with $P=\pm$.
As was argued in \cite{Beisert:2003tq}
this is an indication of integrability. 
Indeed, not only the planar $\alSO(6)$ Hamiltonian
of Minahan and Zarembo \cite{Minahan:2002ve} is 
integrable, but also the \emph{complete} planar 
$\alSL(4|4)$ Hamiltonian \cite{Beisert:2003yb}!
Secondly, we find some overlapping primaries in
\tabref{tab:BabyEnergy},\ref{tab:QuarterEnergy},
clearly their energies do agree. 
What is more, we find that a couple
of energies repeatedly occur. 
These are for example, $\sfrac{3}{4},\sfrac{5}{4},\sfrac{5}{8}, \sfrac{9}{8}$,
but also $\sfrac{1}{8}(5\pm\sqrt{5})$ and $\sfrac{1}{16}(13\pm\sqrt{41})$.
As these states are primaries transforming in different representations,
they cannot be related by $\alSL(4|4)$. 
Of course, these degeneracies could merely be a coincidence of small numbers.
Nevertheless the reappearance of e.g.\ $\sfrac{1}{16}(13\pm\sqrt{41})$ is striking.
This could hint at a further symmetry enhancement
of the planar one-loop Hamiltonian.
It might also turn out to be a consequence of integrability.
Furthermore, one might speculate that it is some
remnant of the broken higher spin symmetry
of the free theory, see e.g.\ \cite{Konshtein:1989yg,Sundborg:2000wp,Sezgin:2001zs,Mikhailov:2002bp}
and references in \cite{Bianchi:2003wx}.

\section{Subsectors}
\label{sec:Sectors}

Especially in view of some recent advances of the 
AdS/CFT correspondence 
\cite{Berenstein:2002jq,Gubser:2002tv,Frolov:2002av}
it has become interesting to determine anomalous dimensions 
of specific operators in $\superN=4$ SYM. 
In principle, the Hamiltonian \eqref{eq:Hfull},\eqref{eq:Hplanar}
provides the answer, but it is hard to apply.
The harmonic action 
\eqref{eq:HamAction},\eqref{eq:HamCoeff},\eqref{eq:HamCoeff00}
is more explicit, but still 
requires a reasonable amount of combinatorics to be applied.
Nevertheless, some operators corresponding to stringy states
have special quantum numbers and often one can restrict to certain 
subsectors of $\superN=4$, see e.g.\ \cite{Beisert:2003xu}.
For instance, in \secref{sec:Baby} we have investigated a 
subsector of $\superN=4$ SYM. Within this subsector
the number of letters as well as the symmetry algebra is reduced.
This reduction of complexity leads to a simplification
of the Hamiltonian \eqref{eq:BabyHam} within the subsector. 
Thus, restricting to subsectors one can efficiently
compute anomalous dimensions.

To construct subsectors, we note that the number of excitations
in the oscillator picture, \eqref{eq:OscExciteab},\eqref{eq:OscExcitec},
naturally puts constraints on the weights of operators. 
Certainly, there cannot be negative excitations.
Furthermore, the oscillators $\oscc^\dagger$ are fermionic. Therefore
there can only be one excitation on each site. In total we find 
twelve bounds
\[\label{eq:SecConstr}
n_{\osca}\geq 0,\quad
n_{\oscb}\geq 0,\quad
n_{\oscc}\geq 0,\quad
L-n_{\oscc}\geq 0.
\]
At the one-loop level all these excitation numbers are conserved.
We can therefore construct `one-loop subsectors' by 
considering operators for which several of these bounds are met. 
In certain cases these subsectors remain closed even at higher loops,
we will refer to these as `closed subsectors'.

\subsection{Closed subsectors}
\label{sec:SecHalfBPS}

Let us demonstrate this procedure for a rather trivial subsector. 
We will consider the subsector of operators with
\[\label{eq:SecHalfBPS}
n_{\osca_{12}}=n_{\oscb^\dagger_{12}}=
n_{\oscc_{12}}=0,
\quad
n_{\oscc_{34}}=L.
\]
In conventional language these operators consist only
of the fields 
\[\label{eq:SecHalfBPSLetter}
Z=\oscc^\dagger_3\oscc^\dagger_4\vac
=\state{Z}=\Phi_{34}=\Phi_{5+i6}=\Phi_{5}+i\Phi_{6}.
\]
These are the half-BPS operators $\Tr Z^L$ and its multi-trace cousins.
The Hamiltonian within this subsector vanishes identically,
as required by protectedness of BPS operators.
Using \eqref{eq:OscExciteab},\eqref{eq:OscExcitec} the constraints \eqref{eq:SecHalfBPS}
allows only the weight 
\[\label{eq:SecHalfBPSWeight}
w=[L;0,0;0,L,0;0,L].
\]
So far we know only that this subsector is closed at one-loop.
To prove closure at higher loops we consider a strictly positive
combination of the bounds
\[\label{eq:SecHalfSum}
n_{\osca_{1}}+
n_{\osca_{2}}+
n_{\oscb_{1}}+
n_{\oscb_{2}}+
n_{\oscc_{1}}+
n_{\oscc_{2}}+
(L-n_{\oscc_{3}})+
(L-n_{\oscc_{4}})=0.
\]
Together with the bounds \eqref{eq:SecConstr} this implies that 
each of the individual terms is zero as in \eqref{eq:SecHalfBPS}. 
Using \eqref{eq:OscExciteab},\eqref{eq:OscExcitec} this combination 
implies 
\[\label{eq:SecHalfSumCharge}
\Delta_0=p.
\]
The label $p$ as well as the \emph{bare} dimension $\Delta_0$ is 
preserved in perturbation theory.%
\footnote{The scaling dimension is obviously not preserved.
The bare dimension is, however, for mixing occurs only
among operators of equal bare dimension.}
Therefore the condition \eqref{eq:SecHalfSumCharge} 
restricts to this subsector at all orders in perturbation theory
which means that this subsector is exactly closed.
We also have that $L=p$, which implies that the length is protected even
at higher loops. Equivalently, the chirality is exactly zero.

Let us investigate all closed subsectors.
For a closed subsector we need to find a positive 
linear combination of the bounds that
is independent of $B$ and $L$.
Put differently, it must be independent of 
$L-B$ and $L+B$. 
The number of excitations $n_{\oscb}$
involves the combination $-B-L$. This can only be
cancelled by $B+L$ in $L-n_{\oscc}$.
Therefore, we can remove oscillators of type $\oscb$
only iff we also fully excite oscillators 
of type $\oscc$.
Equivalently, we can remove oscillators of type $\osca$
only iff we also remove oscillators 
of type $\oscc$.
%
We find the following cases:
\begin{list}{$\bullet$}{\itemsep0pt}
\item
If no oscillator is removed we get the full theory.

\item
If we set $n_{\oscb_2}=0$ and
$n_{\oscc_{i}}=L$, $i=k+1,\ldots,4$,
the remaining symmetry algebra will be
$\alU(1,2|k)$. 
The non-compact form of the algebra is meant to 
indicate that there are infinitely many letters within this subsector.
Furthermore, $B+L$ can be expressed in terms of $p,q_1,q_2$, it is 
therefore conserved.

\item
If we set $n_{\oscb_{1,2}}=0$ 
then we should only set 
$n_{\oscc_4}=L$ in order to get a nontrivial subsector.
The remaining symmetry algebra is $\alSU(2|3)$
and will be discussed in \secref{sec:SecEighthBPS}.
It has conserved $B+L$.

\item
If we set $n_{\osca_{2}}=n_{\oscb_{2}}=n_{\oscc_{i}}=0$, 
$i=1,\ldots k$, and $n_{\oscc_{j}}=L$, $j=l+1,\ldots 4$,
the remaining symmetry algebra will be $\alU(1,1|l-k)$. 
Here, $B$ and $L$ are conserved.
For $k=l=2$ we get the subsector considered in \secref{sec:Baby}.
For $k=l=1,3$ we get a similar subsector
in which the spin of the letters equals $-1$ instead of $-\half$.

\item
If we set $n_{\osca_{2}}=n_{\oscb_{1,2}}=n_{\oscc_{i}}=0$, 
$i=1,\ldots k$, and $n_{\oscc_4}=L$
the remaining symmetry algebra will be $\alU(1|3-k)$.
Here, $B$ and $L$ are conserved.

\item
If we set $n_{\osca_{1,2}}=n_{\oscb_{1,2}}=n_{\oscc_1}=0$, 
and $n_{\oscc_4}=L$
the remaining symmetry algebra will be $\alSU(2)$.
Here, $B$ and $L$ are conserved.
This subsector will be discussed in \secref{sec:SecQuarterBPS}.

\end{list}

\subsection{The nearly quarter-BPS \texorpdfstring{$\alSU(2)$}{su(2)} subsector}
\label{sec:SecQuarterBPS}

In \secref{sec:SecHalfBPS}
we have found that we can express the half-BPS condition 
$\Delta_0=p$ in terms of the excitation numbers of oscillators.
We can do the same for the quarter BPS condition 
\[\label{eq:SecQuarterBPSCharges}
\Delta_0=p+q_1+q_2.
\]
This is equivalent to 
\[\label{eq:SecQuarterSum}
n_{\osca_{1}}+
n_{\osca_{2}}+
n_{\oscb_{1}}+
n_{\oscb_{2}}+
2n_{\oscc_{1}}+
2(L-n_{\oscc_{4}})=0,
\]
which implies that this subsector is closed.
The letters of this subsector are 
\[
Z=\oscc^\dagger_3\oscc^\dagger_4\vac,\quad
\phi=\oscc^\dagger_2\oscc^\dagger_4\vac.
\]
The weights are 
\[
w=[L+\delta \Delta;0,0;n,L-2n,n;0,L],
\]
where $n$ counts the number of $\phi$'s.
The length can be expressed in terms
of $\alSL(4)$ charges, $L=p+q_1+q_2$, it is therefore protected 
within the subsector.
The residual symmetry is $\alSU(2)\times\alU(1)$.
The $\alSU(2)$ factor
transforms $Z$ and $\phi$ in the fundamental representation,
whereas $\alU(1)$ measures the anomalous dimension $\delta\Delta$.
A state with $\delta \Delta=0$ is (at least) quarter-BPS,
a generic state, however, will not be protected. 
Then the weight $w$ is beyond 
the unitarity bounds 
and cannot be primary.
The corresponding primary weight is
(assuming a highest weight state of the residual $\alSU(2)$)
\[
w_0=[L-2+\delta\Delta;0,0;n-2,L-2n,n-2;0,L-2],
\]

This $\alSU(2)$ subsector was studied in \cite{Beisert:2003tq},
where also the two-loop contribution to the dilatation operator was found
\<
\delta D(\gnorm)\eq
-\frac{\gym^2}{8\pi^2}\,\normord{\Tr \comm{\phi}{Z}\comm{\check\phi}{\check Z}}
\nl
-\frac{1}{2} \lrbrk{\frac{\gym^2}{8\pi^4}}^2\Big(
\normord{\Tr \bigcomm{\comm{\phi}{Z}}{\check Z}\bigcomm{\comm{\check\phi}{\check Z}}{Z}}
+\normord{\Tr \bigcomm{\comm{\phi}{Z}}{\check \phi}\bigcomm{\comm{\check\phi}{\check Z}}{\phi}}
\nl
\qquad\qquad\qquad\qquad+\normord{\Tr \bigcomm{\comm{\phi}{Z}}{T^a}\bigcomm{\comm{\check\phi}{\check Z}}{T^a}}
\Big)+\order{\gnorm^6}.
\>
Using this, the first few states and their planar anomalous dimensions were found, 
we list the one-loop part in \tabref{tab:QuarterEnergy}.

\begin{table}\centering
$\begin{array}[t]{|cc|l|}\hline
L&n& \delta \Delta^P\, [\gym^2 N/\pi^2]\\\hline
4&2&\frac{3}{4}^+ \\\hline
5&2&\frac{1}{2}^- \\\hline
6&2&\frac{1}{8}(5\pm \sqrt{5})^+ \\
 &3&\frac{3}{4}^- \\\hline
7&2&\frac{1}{4}^-, \frac{3}{4}^- \\
 &3&\frac{5}{8}^\pm \\\hline
\end{array}
\qquad
\begin{array}[t]{|cc|l|}\hline
L&n& \delta \Delta^P\, [\gym^2 N/\pi^2]\\\hline
8&2&(64x^3-112x^2+56x-7)^+\\
 &3&\frac{1}{2}^\pm, \frac{3}{4}^- \\
 &4&(64x^3-160x^2+116x-25)^+ \\\hline
9&2&\frac{1}{4}(2\pm\sqrt{2})^-,\frac{1}{2}^-\\
 &3&(512x^3-1088x^2+720x-147)^\pm \\
 &4&\frac{5}{8}^\pm,\frac{1}{4}(3\pm \sqrt{3})^- \\\hline
\end{array}$
\caption{
The first few states within the $\alSU(2)$ subsector \cite{Beisert:2003tq}.
The weights of the corresponding primaries are ${[L-2;0,0;n-2,L-2n,n-2;0,L-2]}$.
Cubic polynomials indicate three
states with energies given by the roots of the cubic equation.}
\label{tab:QuarterEnergy}
\end{table}

Furthermore, it was confirmed that the
quarter-BPS operators found in \cite{Ryzhov:2001bp,D'Hoker:2003vf}
are annihilated by the two-loop anomalous dilatation operator 
$\delta D(\gnorm)$. Here, we make the observation that they are
indeed annihilated by the much simpler operator $\mathcal{Q}$ 
(which takes values in the gauge algebra, $\mathcal{Q}=T^a \mathcal{Q}^a$)
\[\label{eq:SecQuarterOp}
\mathcal{Q}=\comm{\check Z}{\check\phi}.
\]
As all terms of $\delta D(\gnorm)$ 
contain this operator, it is clear that 
$\delta D(\gnorm)$ annihilates those quarter-BPS operators. 
We conjecture that all quarter-BPS operators are in the kernel of
$\mathcal{Q}$
\[\label{eq:SecQuarterCondition}
\mathcal{Q}\, \mathcal{O}\indup{1/4-BPS}=0.
\]
This should simplify the search for quarter BPS operators
drastically. We furthermore conjecture that $\mathcal{Q}$ is
an essential part in the first correction to the superboosts,
\[\delta S(\gnorm)\sim \gnorm \,\Tr\Psi \mathcal{Q},\quad
\delta \dot S(\gnorm)\sim \gnorm \,\Tr\bar\Psi \mathcal{Q}.
\]
These generators would join the quarter-BPS multiplet
at primary weight $w$ with three semi-short multiplets
into the long interacting multiplet at primary weight $w_0$. 
In the case of the a quarter-BPS operator, this does not happen
and the multiplet remains short \cite{Dobrev:1985qv,Andrianopoli:1999vr}.

\subsection{The nearly eighth-BPS \texorpdfstring{$\alSU(2|3)$}{su(2|3)} subsector}
\label{sec:SecEighthBPS}

In analogy to \secref{sec:SecQuarterBPS} we investigate
the sector of nearly eighth-BPS operators.
One of the two eighth-BPS conditions is
(the other one would lead to a similar subsector)
\[\label{eq:SecEighthBPSCharges}
\Delta_0=p+\sfrac{1}{2}q_1+\sfrac{3}{2}q_2.
\]
This is equivalent to 
\[\label{eq:SecEighthSum}
n_{\oscb_{1}}+
n_{\oscb_{2}}+
2(L-n_{\oscc_{4}})=0,
\]
which again implies that this subsector is closed.
The letters of this subsector are 
\[
\phi_a=\oscc^\dagger_a\oscc^\dagger_4\vac,\qquad
\psi_\alpha=\osca^\dagger_\alpha\oscc^\dagger_4\vac,\qquad
(a=1,2,3,\,\alpha=1,2)
\]
These transform in the fundamental representation of 
$\alSU(2|3)$. 
We also note that, although $L$ and $B$ are not protected 
in this subsector, the combination $L+B=\Delta_0$ is.
The weights are 
\[
w=[L+B+\delta\Delta;s,0;2B+2L-2p-3q_2,p,q_2;B,L],
\]
where the numbers of individual letters are given by
\[
n_{\phi}=\left(\begin{array}{l}
p+2q_2-2B-L,\\
L-p-q_2,\\
L-q_2,
\end{array}\right)
\qquad
n_{\psi}=
\left(\begin{array}{l}
B+\half s,\\
B-\half s.
\end{array}\right)
\]
This subsector has only finitely many letters. 
The Hamiltonian would therefore have only a few terms 
and it could be applied easily.
Also an investigation of the two-loop contribution to 
the dilatation operator along the lines
of \cite{Beisert:2003tq} seems feasible.
Such an investigation would be very interesting for two reasons.
On the one hand one could see effects of the Konishi anomaly.
The $\order{\gnorm^3}$ (`3/2-loop')
contribution to the dilatation operator
transforms three scalars $\varepsilon^{abc}\phi_a\phi_b\phi_c$ into two
fermions $\varepsilon^{\alpha\beta}\psi_\alpha\psi_\beta$ or vice versa.
In other words it changes the length of the operator.
On the other hand, this sector contains
the subsector of \secref{sec:SecQuarterBPS}. 
In \cite{Beisert:2003tq} 
evidence for the integrability of
the planar two-loop dilatation operator 
was found. Possibly the two-loop integrability 
extends to this subsector and maybe to the full
theory.

The lowest-dimensional eighth-BPS operator 
is expected to be a triple-trace operator
with weight 
\[
w=[6;0,0;0,0,4;0,6].
\]
We find this operator, it is
\<\label{eq:SecEighthOp}
\Op\indup{1/8-BPS}\eq
\varepsilon^{abc}\varepsilon^{def}\big[
N(N^2-3)\Tr \phi_a \phi_d \Tr \phi_b \phi_e \Tr \phi_c \phi_f
\nl
\qquad\qquad+6(N^2-1)\Tr \phi_a \phi_d \Tr \phi_b \phi_c \phi_e \phi_f
\nl
\qquad\qquad-12N\Tr \phi_a \phi_b \phi_c \phi_d \phi_e \phi_f
\nl
\qquad\qquad+8N\Tr \phi_a \phi_d \phi_b \phi_e \phi_c \phi_f
\nl
\qquad\qquad+4\Tr \phi_a \phi_b \phi_c \Tr \phi_d \phi_e \phi_f\big].
\>
It is annihilated by the generalisation of 
\eqref{eq:SecQuarterOp}
\[
\mathcal{Q}_{a}=\varepsilon_{abc}\comm{\check\phi^b}{\check\phi^c}
\]
which implies that it is protected at one-loop.
It is also annihilated by the operator
\[
D_3\sim\varepsilon_{abc}\varepsilon^{\alpha\beta}\Tr \psi_\alpha\comm{\check\phi^a}{\comm{\check\phi^b}{\comm{\check\phi^c}{\psi_\beta}}}.
\]
An investigation of diagrams shows that the relevant
$3/2$-loop contribution to the dilatation generator
should be proportional to this operator. 
This supports the claim that the operator
\eqref{eq:SecEighthOp} is eighth-BPS.

\subsection{The \texorpdfstring{$\alSO(6)$}{so(6)} subsector}
\label{sec:SecSO6}

This is the subsector where the scalars are the only letters
\[\label{eq:SecSO6}
n_{\osca_{12}}=n_{\oscb_{12}}=0.
\]
The allowed weights are
\[
w=[L;0,0;q_1,p,q_2;0,L].
\]
The residual symmetry is $\alSL(4)=\alSO(6)$.
This subsector is a one-loop subsector. All bounds
$n_{\osca},n_{\oscb}$ 
have a negative coefficient of $L$. Therefore
all positive combinations of bounds involve $L$,
which is broken at higher loops.

We will investigate the non-planar Hamiltonian
in this sector. Two scalars $\Phi_p,\Phi_q$
can be symmetrised in three different ways,
symmetric-traceless, antisymmetric and singlet.
These correspond to the modules $V_0,V_1,V_2$, respectively.
The projectors to these representations are
\<
(P_0)_{mn}^{pq}\eq\half\delta_{m}^{p}\delta_{n}^{q}+\half\delta_{n}^{p}\delta_{m}^{q}-\sfrac{1}{6}\delta_{mn}\delta^{pq},
\nln
(P_1)_{mn}^{pq}\eq\half \delta_{m}^{p}\delta_{n}^{q}-\half \delta_{n}^{p}\delta_{m}^{q},
\nln
(P_2)_{mn}^{pq}\eq\sfrac{1}{6}\delta_{mn}\delta^{pq},
\>
The coefficient \eqref{eq:D2Coeff} of the Hamiltonian 
using the harmonic eigenvalues \eqref{eq:HarmCoeff} is
\[
C_{mn}^{pq}=
0\cdot (P_0)_{mn}^{pq}+
1\cdot(\half \delta_{m}^{p}\delta_{n}^{q}-\half \delta_{n}^{p}\delta_{m}^{q})
+\sfrac{3}{2}\cdot(\sfrac{1}{6}\delta_{mn}\delta^{pq})
\]
We substitute this in the 
Hamiltonian \eqref{eq:Hreduce} and get
\<
H\eq
N^{-1} \bigbrk{-\half\,\normord{\Tr\comm{\Phi_m}{\check \Phi^m}\comm{\Phi_n}{\check \Phi^n}}
+\half\,\normord{\Tr\comm{\Phi_m}{\check \Phi^n}\comm{\Phi_n}{\check \Phi^m}}
-\sfrac{1}{4}\, \normord{\Tr\comm{\Phi_m}{\check \Phi^n}\comm{\Phi_m}{\check \Phi^n}}}
\nle
N^{-1} \bigbrk{
-\half\,\normord{\Tr\comm{\Phi_m}{\Phi_n}\comm{\check \Phi^m}{\check \Phi^n}}
-\sfrac{1}{4}\, \normord{\Tr\comm{\Phi_m}{\check \Phi^n}\comm{\Phi_m}{\check \Phi^n}}}.
\>
After multiplication with $\gym^2 N/8\pi^2$ this is exactly the effective vertex 
found in \cite{Beisert:2002bb} and yields the dilatation generators 
in \cite{Minahan:2002ve,Beisert:2002ff,Beisert:2003tq}.

\subsection{The \texorpdfstring{$\alSO(4,2)$}{so(4,2)} subsector}
\label{sec:SecSO42}

One might choose to set
\[
n_{\oscc_{12}}=0,\quad
n_{\oscc_{34}}=L,
\]
or, more conveniently,
$n_{\oscc_{12}}=n_{\oscd_{12}}=0$ using the primary
vacuum.
This is a generalisation of the subsector discussed in 
\secref{sec:Baby}, which is, however, closed only at one-loop.
In this subsector the letters are the scalars $Z$ with any number of 
the four spacetime derivatives acting
\[
(\osca^\dagger_1)^{k_1}
(\osca^\dagger_2)^{k_2}
(\oscb^\dagger_1)^{l_1}
(\oscb^\dagger_2)^{k_1+k_2-l_1}
\state{Z}.
\]
The weight of a state is given by
\[
w=[\Delta_0;s_1,s_2;0,L,0;0,L]
\]
where the total numbers of excitations are
\[
k_1=\half\Delta_0-\half L+\half s_1,\quad
k_2=\half\Delta_0-\half L-\half s_1,\quad
l_1=\half\Delta_0-\half L+\half s_2.
\]
This sector might be useful to investigate 
semiclassical strings 
spinning on one circle in $S^5$ and two circles in $AdS^5$.

\subsection{The \texorpdfstring{$\alSU(2|4)$}{su(2|4)} one-loop BMN matrix model}

The BMN matrix model \cite{Berenstein:2002jq}
in the one-loop approximation \cite{Kim:2003rz}
is obtained by setting
\[\label{eq:SecBMN}
n_{\oscb_{12}}=0.
\]
Again, this yields only a one-loop subsector.
The letters $W''_A$ of the matrix model are
\[
\oscc^\dagger_a\oscc^\dagger_b\vac,\quad
\osca^\dagger_\alpha\oscc^\dagger_b\vac,\quad
\osca^\dagger_\alpha\osca^\dagger_\beta\vac
\]
corresponding to the $\alSO(6)$ vectors, fermions and $\alSO(3)$ vectors.
The residual symmetry is $\alSU(2|4)$.
The multiplet of letters $V''\indup{F}$ is given by the primary weight
\[
V''\indup{F}=[q_1,p,q_2]^{\Delta_0}_{s}=[0,1,0]_0^1.
\]
The irreducible modules of two multiplets of letters are
\[
V''\indup{F}\times V''\indup{F}=V''_0+V''_1+V''_2
\]
with 
\[V''_0=[0,2,0]_0^2,\quad
V''_1=[1,0,1]_0^2,\quad
V''_2=[0,0,0]_0^2.\]
In $\alSL(2)\times\alSL(4)$ these modules split into
\<
V''_0\eq  [0,2,0]_0^{2}
         +[0,1,1]_1^{2.5}
         +[0,1,0]_2^{3}+[0,0,2]_0^{3}
         +[0,0,1]_1^{3.5}
         +[0,0,0]_0^{4},
\nln
V''_1\eq  [1,0,1]_0^2
         +[0,1,1]_1^{2.5}+[1,0,0]_1^{2.5}
         +[0,0,2]_2^{3}  +[0,1,0]_2^{3}  +[0,1,0]_0^{3}
\nl
         +[0,0,1]_3^{3.5}+[0,0,1]_1^{3.5}
         +[0,0,0]_2^{4},
\nln
V''_2\eq  [0,0,0]_0^{2}
         +[1,0,0]_1^{2.5}
         +[0,1,0]_2^{3}
         +[0,0,1]_3^{3.5}
         +[0,0,0]_4^{4}.
\>
In \cite{Kim:2003rz} it was found that the one-loop spectrum for 
the matrix model in the $\alSO(6)$ vector sector matches the 
$\alSO(6)$ sector of $\superN=4$ SYM.
The complete set of modules $V''_j$ of the full 
$\alSU(2|4)$ matrix model is already realised in the
$\alSO(6)$ subsector. Therefore the
$\alSO(6)$ subsector lifts uniquely to the
full $\alSU(2|4)$ matrix model.
Thus the restriction of the one-loop $\superN=4$ SYM dilatation operator 
to this subsector agrees with 
the complete matrix model Hamiltonian.
In other words the matrix model Hamiltonian $H''=MD''/2$ is given by
\[\label{eq:HBMNfull}
D''(1/M)=D''_0-\frac{4}{M^3}
\lrbrk{(P''_1)_{CD}^{AB}+\sfrac{3}{2}(P''_2)_{CD}^{AB}}
\normord{\Tr \comm{W''_A}{\check W^{\prime\prime\,C}}\comm{W''_B}{\check W^{\prime\prime\,D}}}
+\order{M^{-4}},
\]
where the precise form of the projectors $P''_{1,2}$ remains to be evaluated.
Alternatively, 
the Hamiltonian density could be determined using
\eqref{eq:HamAction},\eqref{eq:HamCoeff},\eqref{eq:HamCoeff00}.

\section{Outlook}
\label{sec:Concl}


Here we have constructed the 
highly intricate, first radiative corrections
to the (trivial) classical dilatation operator.
One might also investigate radiative corrections to the other 
generators of $\alPSL(4|4)$. 
The generators that receive corrections are the momenta 
and boosts, $P,K,Q,\dot Q,S,\dot S$.
The closure of the algebra, see \appref{sec:alg}, should put 
tight constraints on these as well as on the dilatation generator.
In fact, these might determine the dilatation operator at one-loop
or even higher! 
Let us demonstrate this using the related 
issue of multiplet splitting at the 
unitarity bounds \cite{Andrianopoli:1999vr}:
We have constructed $H$ as an invariant 
operator under the classical $\alPSL(4|4)$.
In the classical $\alPSL(4|4)$, 
long multiplets at the unitarity bounds split up. 
In the interacting theory
these multiplets must rejoin. This is, however, only possible
if all submultiplets have degenerate anomalous dimensions. 
For instance, we find the following three energies for
the Konishi submultiplets
\[
E=4C_2=6C_1=6C_1-2C_0,
\]
where we have left the independent coefficients
\eqref{eq:D2Coeff} unfixed.
This consistency requirement determines $C_0,C_1,C_2$
up to an overall constant.
A similar argument was used in 
\cite{Anselmi:1997mq,Anselmi:1998ms} to determine some anomalous dimensions.
One might hope that an investigation of 
the twist-two operators \eqref{eq:Pomeron}
might constrain \emph{all} independent coefficients to $C_j=c\, h(j)$.
If this works out, all one-loop anomalous dimensions can be obtained
purely algebraically without evaluating a single Feynman diagram
up to one overall constant!
This constant can finally be fixed
by a different consistency argument \cite{Anselmi:1997mq,Anselmi:1998ms},
which merely requires computing the quotient of two tree-level diagrams.
In that spirit it would be very interesting to find out if
these consistency arguments or, more generally, the closure of the interacting algebra, 
can be used to fix the higher-loop contributions to the 
dilatation generator as well.
This is not inconceivable, given that the $\superN=4$ action is known 
to be unique.

It would be great to obtain higher-loop contributions to
the dilatation generator.
If the one-loop contribution turns out 
to be completely fixed by symmetry,
only higher-loop anomalous dimensions
could provide truly dynamical information about $\superN=4$ SYM
and the dynamical AdS/CFT correspondence.
Even better, one might find that \emph{all} radiative
corrections are somehow fixed by symmetry and thus
kinematical. 
Also the question of higher-loop integrability raised in 
\cite{Beisert:2003tq} could be addressed. 
As a starting point, one might restrict to the $\alSU(2|3)$ subsector of 
\secref{sec:SecEighthBPS} or the non-compact $\alSL(2)$ subsector
of \secref{sec:Baby} to simplify the computations. 
Eventually a treatment of the complete dilatation generator at two-loops
would be desirable. This might be feasible using computer algebra packages 
developed for higher-loop calculations within QCD.

The idea of investigating the dilatation operator can be generalised to 
a wider range of QFT's. 
For instance, a few theories with $\superN=2$ supersymmetry are 
conformal at the quantum level. For these the determination 
of the dilatation generator might shed some light on
holographic dualities
correspondence away from the well-studied case of ${AdS_5\times S^5}$.
Even in a QFT without conformal invariance the techniques 
developed in \cite{Beisert:2002ff,Beisert:2003tq}
can be used to investigate 
logarithmic corrections 
to two-point functions and scattering amplitudes 
in a systematic way.
In particular in QCD at large $N\indup{c}$ 
and deep inelastic scattering, 
similar techniques are at use (see e.g.\ \cite{Faddeev:1995zg,Braun:1998id,Belitsky:2003ys}).
There, following pioneering work of Lipatov \cite{Lipatov:1994yb}, 
methods of integrability have also had much impact.

An intriguing feature of 
\tabref{tab:BabyEnergy},\ref{tab:anotab},\ref{tab:QuarterEnergy}
is that several states have degenerate energies, although they are
not related by $\alSL(4|4)$ symmetry,
see the discussion at the end of 
\secref{sec:OscEx}.
The occurrence of parity pairs provides some evidence
for integrability.
Integrability of the full planar theory at one-loop
and the corresponding Bethe ansatz equations 
are presented in \cite{Beisert:2003yb},
reproducing the anomalous dimensions in the above tables.
Apart from these, we find more examples of degeneracies.
Could this be the result of a further symmetry enhancement 
due to or beyond integrability?

\phantomsection
\addcontentsline{toc}{subsection}{Acknowledgements}
\subsection*{Acknowledgements}

The author benefited from interesting discussions with
Thomas Klose,
Stefano Kovacs,
Charlotte Kristjansen,
Pedro Liendo,
Joe Minahan, 
Francisco Morales, 
Jan Plefka
and
Kostya Zarembo.
Furthermore, I would like to thank Gleb Arutyunov for
sharing his insights in superconformal symmetry and
for reading the manuscript.
In particular I am grateful to Matthias Staudacher,
for constant interest, general support, and helpful remarks 
regarding this project. 
I would like to thank the 
Laboratori Nazionali di Frascati and the 
Department of Theoretical Physics in Uppsala
for their hospitality during the early stages of this project. 
Ich danke der \emph{Studienstiftung des
deutschen Volkes} f\"ur die Unterst\"utzung durch ein 
Promotions\-f\"orderungsstipendium.

\appendix

\section{The algebra \texorpdfstring{$\alGL(4|4)$}{gl(4|4)}}
\label{sec:alg}

The algebra $\alGL(4|4)$ consists of the generators
$J=(Q,S,\dot Q,\dot S,P,K,L,\dot L,R,D,C,B)$.
These are the (super)translations $Q,\dot Q,P$, 
the (super)boosts $S,\dot S,K$,
the $\alSL(2)\times\alSL(2)$ rotations $L,\dot L$,
the $\alSL(4)$ rotations $R$
as well as the dilatation generator $D$, central charge $C$ and
chirality $B$. 

Under the rotations $L,\dot L,R$, the indices of any generator $J$
transform canonically according to 
\[\label{eq:gl44rot}
\arraycolsep0pt
\begin{array}{rclcrcl}
\comm{L^\alpha{}_\beta}{J_\gamma}\eq
\delta^\alpha_\gamma J_\beta
-\half \delta^\alpha_\beta J_\gamma,
&\quad&
\comm{L^\alpha{}_\beta}{J^\gamma}\eq
-\delta^\gamma_\beta J^\alpha
+\half \delta^\alpha_\beta J^\gamma,
\\
\comm{\dot L^{\dot\alpha}{}_{\dot\beta}}{J_{\dot\gamma}}\eq
 \delta^{\dot\alpha}_{\dot\gamma} J_{\dot\beta}
-\half \delta^{\dot\alpha}_{\dot\beta} J_{\dot\gamma},
&&
\comm{\dot L^{\dot\alpha}{}_{\dot\beta}}{J^{\dot\gamma}}\eq
-\delta_{\dot\beta}^{\dot\gamma} J^{\dot\alpha}
+\half \delta^{\dot\alpha}_{\dot\beta} J^{\dot\gamma},
\\
\comm{R^a{}_b}{J_c}\eq
\delta^a_c J_b
-\quarter \delta^a_b J_c,
&&
\comm{R^a{}_b}{J^c}\eq
-\delta_b^c J^a
+\quarter \delta^a_b J^c.
\end{array}\]
The charges $D,C,B$ of the generators are given by
\[\label{eq:gl44charge}
\comm{D}{J}=\dim(J)\, J,\qquad \comm{C}{J}=0,\qquad \comm{B}{J}= \mathrm{chi}(J)\, J
\]
with non-vanishing dimensions
\[\label{eq:gl44dim}
\dim(P)=-\dim(K)=1,\quad \dim(Q)=\dim(\dot Q)=-\dim(S)=-\dim(\dot S)=\half
\]
and non-vanishing chiralities
\[\label{eq:gl44chi}
\mathrm{chi}(Q)=-\mathrm{chi}(\dot Q)=-\mathrm{chi}(S)=\mathrm{chi}(\dot S)=\half.
\]
The translations and boosts 
commuting into themselves are given by
\[\label{eq:gl44mommom}
\arraycolsep0pt
\begin{array}{rclcrcl}
\comm{S^\alpha{}_a}{P_{\beta\dot\gamma}}\eq 
   \delta^\alpha_\beta \dot Q_{\dot\gamma a},
&\qquad&
\comm{K^{\alpha\dot\beta}}{\dot Q_{\dot\gamma c}}\eq
  \delta^{\dot\beta}_{\dot\gamma} S^\alpha{}_c,
\\[3pt]
\comm{\dot S^{\dot\alpha a}}{P_{\beta\dot\gamma}}\eq 
  \delta^{\dot\alpha}_{\dot\gamma} Q^a{}_{\beta},
&&
\comm{K^{\alpha\dot\beta}}{Q^c{}_\gamma}\eq
  \delta^\alpha_\gamma \dot S^{\dot\beta c},
\\[3pt]
\acomm{\dot Q_{\dot\alpha a}}{Q^b{}_\beta}\eq 
  \delta_a^b P_{\beta\dot\alpha},
&&
\acomm{\dot S^{\dot\alpha a}}{S^\beta{}_b}\eq 
  \delta^a_b K^{\beta\dot\alpha},
\end{array}
\]
while the translations and boosts commuting into rotations are given by
\<\label{eq:gl44momrot}
\comm{K^{\alpha \dot\beta}}{P_{\gamma\dot\delta}}\eq 
  \delta_{\dot\delta}^{\dot\beta} L^{\alpha}{}_{\gamma}
  +\delta_\gamma^\alpha \dot L^{\dot\beta}{}_{\dot\delta}
  +\delta_\gamma^\alpha\delta_{\dot\delta}^{\dot\beta} D
,
\nln
\acomm{S^\alpha{}_a}{Q^b{}_\beta}\eq
  \delta^b_a L^\alpha{}_\beta
  +\delta_\beta^\alpha R^b{}_a
  +\half \delta_a^b \delta_\beta^\alpha (D-C),
\nln
\acomm{\dot S^{\dot \alpha a}}{\dot Q_{\dot\beta b}}\eq
  \delta^a_b \dot L^{\dot\alpha}{}_{\dot\beta}
  -\delta_{\dot\beta}^{\dot\alpha} R^a{}_b
  +\half \delta^a_b \delta_{\dot\beta}^{\dot\alpha} (D+C).
\>
The chirality $B$ never appears on the right hand side,
it can be dropped. Furthermore, the central charge can be 
set to zero. The resulting algebra is $\alPSL(4|4)$.

The quadratic Casimir of $\alGL(4|4)$ is
\<\label{eq:gl44Casimir}
J^2\eq 
\half D^2
+\half L^\gamma{}_\delta L^\delta{}_\gamma
+\half \dot L^{\dot\gamma}{}_{\dot\delta} \dot L^{\dot\delta}{}_{\dot\gamma}
-\half R^c{}_d R^d{}_c
\nl
-\half \comm{Q^c{}_\gamma}{S^\gamma{}_c}
-\half \comm{\dot Q_{\dot\gamma c}}{\dot S^{\dot\gamma c}}
-\half \acomm{P_{\gamma\dot\delta}}{K^{\gamma\dot\delta}}
-BC.
\>
For $\alPSL(4|4)$ the last term $BC$ is absent.

\section{Calculation of diagrams}
\label{sec:diagrams}

The two-point function of operators with vector indices 
is restricted by conformal symmetry to
\[
\bigvev{\Op_{a,\mu_1\ldots\mu_m}(x)\,\Op_{b,\nu_1\ldots\nu_n}(y)}
=\frac{\delta_{ab}(N_a+\gnorm^2 N'_a)}{(x-y)^{2\Delta_0+2\gnorm^2\delta\Delta_a}}\,
J_{\mu_{i_1}\nu_{i_1}} J_{\mu_{i_2}\nu_{i_2}} \delta_{\mu_{i_3}\mu_{i_3}} \delta_{\nu_{i_4}\nu_{i_4}}\ldots\,.
\]
The symbols 
\[J_{\mu\nu}=\delta_{\mu\nu}-2\frac{(x-y)_\mu (x-y)_\nu}{(x-y)^2}\]
relate conformal tangent spaces at different space-time points.
The vectors $x-y$ with open indices 
do not carry essential information, they can be discarded
and reconstructed later by replacing $\delta_{\mu\nu}$ by
$J_{\mu\nu}$ where appropriate.
Also the corrections to the normalisation constants $N'$ are irrelevant
to the anomalous dimension at one-loop.
Before diagonalisation we thus expect the correlator to have the 
following structure 
\[
\frac{N_{ac}(\delta^c_b+\gnorm^2\delta\Delta^c{}_b\log |x-y|^{-2})}{(x-y)^{2\Delta_0}}\,
\delta_{\mu_{i_1}\nu_{i_1}} \delta_{\mu_{i_2}\nu_{i_2}} \delta_{\mu_{i_3}\mu_{i_3}} \delta_{\nu_{i_4}\nu_{i_4}}\ldots
\]
We use the letters
\[
Z_k=(\osca^\dagger)^k\vac=\frac{1}{k!}(\cder_{1}+i\cder_{2})^k (\Phi_{5}+i\Phi_{6}).
\]
The expansion of the covariant derivative $\cder=\partial-i\gnorm \fldA$
can move one `bulk' vertex to the position of the field
and become a `boundary' vertex. At one-loop there can only
be one boundary vertex, the contribution from two boundary vertices
has no logarithmic behaviour. The relevant part of the letter $Z_k$ 
is thus
\[
Z_k=\frac{1}{k!}\,\partial^k \Phi
-i\gnorm \sum_{j=1}^k \frac{1}{j!(k-j)!}\,\comm{\partial^{j-1} \fldA}{\partial^{k-j}\Phi}.
\]

We go ahead and calculate 
the relevant parts of the correlator
\[\bigvev{Z_k(x_1)\, Z_{n-k}(x_2)\, \bar Z_{k'}(x_3)\, \bar Z_{n-k'}(x_4)}.\]
We restrict to the
planar sector and use point splitting
regularisation.
The matrix of anomalous dimensions $\delta \Delta$
is generated by the Hamiltonian, we which we assume the generic form
\[
H_{12} \,(\osca^\dagger_1)^k (\osca^\dagger_2)^{n-k}\state{00}=
\sum_{k'=0}^n c_{n,k,k'}\,(\osca^\dagger_1)^{k'} (\osca^\dagger_2)^{n-k'}\state{00}.
\]
Each diagram contributes a set of constants $c_{n,k,k'}$, which
we will list below. We list the contributions separately, they 
can be reused for a calculation within a different theory.

The diagrams \figref{fig:diagrams}a
with an intermediate gluon 
(up to terms where one scalar line has been collapsed to a point
by the equations of motion)
yield the vertex 
\[
\frac{\gym^2N}{32\pi^2}\,
\frac{\partial_1^{k}\partial_2^{n-k}\partial_3^{k'}\partial_4^{n-k'}}{k!(n-k)!k'!(n-k')!}\,
\frac{(r-s)\,\Phi(r,s)}{x_{13}^2x_{24}^2}\,,
\]
where $r,s$ are the conformal cross ratios
\[
r=\frac{x_{12}^2 x_{34}^2}{x_{13}^2 x_{24}^2}\,,\quad
s=\frac{x_{14}^2 x_{23}^2}{x_{13}^2 x_{24}^2}\,.
\]
We use the expansion \cite{Arutyunov:2000ku}
of $\Phi$ in the limit $x_2\to x_1$, $x_4\to x_3$ ($r\to 0, s\to 1$)
\[
\Phi(r,s)=
-\sum_{n,m=0}^\infty\frac{(n+m)!^2}{ m! (1+2n+m)!}\, r^n (1-s)^m \log r+\ldots
\]
and obtain the coefficient
\[
c_{n,k,k'}=
\frac{1}{2(n+1)}+
\delta_{k=k'}\bigbrk{\half \harm{k}+\half \harm{n-k}}-\frac{\delta_{k\neq k'}}{2|k-k'|}\,.
\]
Diagram \figref{fig:diagrams}a with a
four-point interaction is given by
\[
\frac{\gym^2N}{32\pi^2}\,
\frac{\partial_1^{k}\partial_2^{n-k}\partial_3^{k'}\partial_4^{n-k'}}{k!(n-k)!k'!(n-k')!}\,
\frac{\Phi(r,s)}{x_{13}^2 x_{24}^2}
\]
and the corresponding coefficient is
\[
c_{n,k,k'}=
-\frac{1}{2(n+1)}\,.
\]
Diagrams \figref{fig:diagrams}a
with an intermediate gluon where
one scalar line has been collapsed to a point by the equations of motion 
is given by
\[
\frac{\gym^2N}{32\pi^2}\,
\frac{\partial_1^{k}\partial_2^{n-k}\partial_3^{k'}\partial_4^{n-k'}}{k!(n-k)!k'!(n-k')!}\,
\frac{(s'-r')\Phi(r',s')}{x_{13}^2x_{24}^2}
+\mbox{3 perm.}
\]
with
\[
r'=\frac{x_{12}^2}{x_{24}^2},\quad
s'=\frac{x_{14}^2}{x_{24}^2}
\]
and the corresponding coefficient is
\<
c_{n,k,k'}\eq
\delta_{k=k'}\bigbrk{-\half \harm{k+1}-\half \harm{n-k+1}}
+\frac{\delta_{k\neq k'}}{2|k-k'|}
\nl
-\frac{\delta_{k> k'}}{4(k+1)}
-\frac{\delta_{k< k'}}{4(k'+1)}
-\frac{\delta_{k< k'}}{4(n-k+1)}
-\frac{\delta_{k> k'}}{4(n-k'+1)}\,.
\>
Diagrams \figref{fig:diagrams}a$'$ yields
\[
\frac{\gym^2N}{32\pi^2}\,\sum_{j=1}^k
\lrbrk{\frac{\partial_1^{k-j}\partial_3^{k'}}{(k-j)!k'}\,\frac{1}{x_{13}^2}}
\lrbrk{\frac{\partial_1^{j-1}(\partial_{1}+2\partial_{2})\partial_2^{n-k}\partial_4^{n-k'}}{j!(n-k)!(n-k')!}\,
\frac{\Phi(r',s')}{x_{24}^2}}
+\mbox{3 perm.}
\]
and the corresponding coefficient is
\[
c_{n,k,k'}=
-\frac{\delta_{k\neq k'}}{|k-k'|}
+\frac{\delta_{k> k'}}{4(k+1)}
+\frac{\delta_{k< k'}}{4(k'+1)}
+\frac{\delta_{k< k'}}{4(n-k+1)}
+\frac{\delta_{k> k'}}{4(n-k'+1)}\,.
\]
The contributions from diagrams \figref{fig:diagrams}c with an intermediate gluon,
with intermediate fermions and the contributions from \figref{fig:diagrams}c',
respectively, are given by 
\<
c_{n,k,k'}\eq -\delta_{k=k'},
\nln
c_{n,k,k'}\eq 2\delta_{k=k'},
\nln
c_{n,k,k'}\eq
\delta_{k=k'}
\bigbrk{\half \harm{k}+\half \harm{n-k}+\half \harm{k+1}+\half \harm{n-k+1}-1}.
\>
The sum of all coefficients is
\[
c_{n,k,k'}=
\delta_{k=k'}\bigbrk{\harm{k}+\harm{n-k}}
-\frac{\delta_{k\neq k'}}{|k-k'|}\,.
\]


\phantomsection
\addcontentsline{toc}{section}{\refname}
\bibliography{collectreduce5}
\bibliographystyle{nb}

\end{document}